\documentclass[a4paper,11pt]{article}

\usepackage{jheppub} 

\usepackage[T1]{fontenc} 

\usepackage{graphicx}
\usepackage{dcolumn}
\usepackage{bm}

\usepackage{relsize}
 \usepackage{multirow,graphics}
 \usepackage{amstext}
 \usepackage{amssymb}
 \usepackage{amsmath}
 \usepackage{graphicx}
 \usepackage{color}
 \usepackage{dsfont} 
 \usepackage{tikz}
\usepackage{epsfig}
\usepackage{float}

\newcommand{\CN}{{\cal N}}

\newcommand{\p}{\partial}
\newcommand{\Tr}{{\text{Tr}}}

\newcommand \ket [1] {|{#1}\rangle}

\newcommand{\pa}{\partial}

\DeclareMathOperator{\idop}{\mathds{1}}
\newcommand{\sfrac}[2]{{\textstyle\frac{#1}{#2}}}
\newcommand{\half}{\sfrac{1}{2}}

\newcommand{\quarter}{\sfrac{1}{4}}
\newcommand{\gen}[1]{#1}
\newcommand{\comm}[2]{[#1,#2]}
\newcommand{\levo}{\widehat J}
\newcommand{\levz}{J}
\newcommand{\alg}[1]{\mathfrak{#1}}
\newcommand{\Jexp}{\mathcal{J}}

\renewcommand{\a}{\alpha}
\newcommand{\da}{{\dot{\alpha}}}
\newcommand{\db}{{\dot{\beta}}}
\renewcommand{\b}{\beta}
\newcommand{\g}{\gamma}

\newcommand{\dg}{{\dot{\gamma}}}
\newcommand{\la}{\lambda}
\newcommand{\tla}{\tilde\lambda}

\newcommand{\genosr}[1]{\mathfrak{#1}}

\makeatletter
\newlength{\apb@width}
\newcommand{\autoparbox}[2][c]{\settowidth{\apb@width}{#2}\parbox[#1]{\apb@width}{#2}}
\newcommand{\includegraphicsbox}[2][]{\autoparbox{\includegraphics[#1]{#2}}}
\makeatother

\title{\boldmath Yangian Symmetry for Bi-Scalar Loop Amplitudes}

\preprint{HU-EP-17/09,\quad  LPTENS-17/**,\quad MITP/17-022}

\author{Dmitry  Chicherin$^a$,}  
\author{Vladimir Kazakov$^{b}$,}
\author{Florian Loebbert$^c$,}  
\author{Dennis M\"uller$^c$,}  
\author{De-liang Zhong$^{b}$}

\affiliation[a]{PRISMA Cluster of Excellence, Johannes Gutenberg University, 55099 Mainz, Germany}
\affiliation[b]{Laboratoire de Physique Th\'eorique, D\'epartement de Physique de l'ENS, Ecole Normale Sup\'erieure, PSL Research University, Sorbonne Universit\'es,  UPMC Univ. Paris 06, CNRS, 75005 Paris, France}
\affiliation[c]{Institut f\"ur Physik, Humboldt-Universi\"at zu Berlin,
Zum Gro{\ss}en Windkanal 6, 12489 Berlin, Germany}

\emailAdd{chicherin@uni-mainz.de}
\emailAdd{kazakov@physique.ens.fr}
\emailAdd{loebbert@physik.hu-berlin.de}
\emailAdd{dmueller@physik.hu-berlin.de}
\emailAdd{zdlzdlzdl@gmail.com}

\abstract{
We establish an all-loop  conformal Yangian symmetry for the full set of planar amplitudes  in the recently proposed integrable bi-scalar field  theory in four dimensions. This chiral theory is a particular  double scaling limit of \(\gamma\)-twisted  weakly coupled $\mathcal{N}=4$ SYM theory. Each amplitude with a certain order of scalar particles is given by a single fishnet Feynman graph of disc topology cut out of a regular square lattice. The Yangian can be realized by the action of a product of  Lax operators with a specific sequence of inhomogeneity parameters on the boundary of the disc. Based on this observation, the Yangian generators of level one for generic bi-scalar amplitudes are explicitly constructed. Finally, we comment on the relation to the dual conformal symmetry of these scattering amplitudes. \\

\bigskip
}  

\dedicated{This paper is dedicated to the memory of L.D.Faddeev.}

\usepackage{amsmath}

\begin{document} 

\maketitle
\flushbottom

\section{Introduction}\label{sec:intro}

The  quantum  integrability of  planar $\mathcal{N}=4$ super Yang--Mills (SYM) theory \cite{Beisert:2010jr} furnishes an important instrument for the study of numerous physical quantities within this conformal field theory (CFT). Among these are the spectrum of anomalous dimensions of local operators --- especially with the discovery of the quantum spectral curve (QSC) \cite{Gromov2014a,Gromov:2014caa,Kazakov:2015efa} ---  conformal structure constants \cite{Basso:2015zoa}, multi-point correlators \cite{Fleury:2016ykk}, cusped Wilson loops, the quark-antiquark potential \cite{Gromov:2016rrp}, as well as planar scattering amplitudes. 

Planar amplitudes --- some of the most important physical quantities --- have been extensively studied in recent years; see e.g.\ \cite{2011JPhA...44a0101R,Elvang:2013cua} for reviews. 
In order to use integrability for their computation, a few very promising ideas have been proposed, such as the  pentagon OPE of \cite{Basso:2013vsa} for instance. However, a systematic and efficient integrability toolbox for amplitudes is still missing. The symmetry that underlies the integrability of the planar S-matrix is the so-called dual conformal \cite{Drummond:2008vq} or Yangian symmetry \cite{Drummond:2009fd}. The Yangian symmetry has been well established as a symmetry of tree-level scattering amplitudes, but its generalization to higher loop orders is hindered by IR singularities which, being regularized, destroy this  symmetry \cite{Beisert:2010gn}.%
\footnote{See \cite{CaronHuot:2011kk} for an approach towards Yangian symmetry of the finite BDS-subtracted scattering matrix of $\mathcal{N}=4$ SYM theory. In the Grassmannians 
 formulation of scattering amplitudes, the integrand has manifest Yangian symmetry, cf.\ \cite{ArkaniHamed:2012nw}.}

However, as we will show in this paper, an all-loop Yangian symmetry can be constructed for the case of scalar amplitudes within the recently proposed, by one of the authors and \"O.~G\"urdogan in \cite{Gurdogan:2015csr}, double scaling limit of \(\gamma\)-deformed \(\CN=4\) SYM theory. This limit  is obtained by taking the 't~Hooft coupling $g\to0$, the \(\gamma\)-deformation parameters $e^{-i\gamma_j/2}\to\infty$, while keeping the new couplings \(\xi_j=ge^{-i\gamma_j/2}\)(\(j=1,2,3)\) fixed. In the particular case of a single non-zero effective coupling \(\xi\equiv \xi_3\), this limit gives rise to the so-called bi-scalar \(\chi\)FT\(_4\)  with the four-dimensional Lagrangian   
 \begin{align}
    \label{Lthree}
    {\cal L}_{\phi}= \frac{N_c}{2}\Tr\,\,
    \left(\p^\mu\phi^\dagger_1 \p_\mu\phi_1+\p^\mu\phi^\dagger_2 \p_\mu\phi_2+2\xi^2\,\phi_1^\dagger \phi_2^\dagger \phi_1\phi_2\right)\,.\hfill\,
  \end{align}  
 Here $\phi_1$ and $\phi_2$ are two complex scalar matrix fields in the adjoint representation of \(SU(N_c) \). The absence of the complex conjugate term \(\Tr\,\big(\phi_2^\dagger \phi_1^\dagger \phi_2\phi_1\big)\) makes the model non-unitary, which implies specific chiral properties with respect to its  flavor structure.  
 In the 't~Hooft limit \(N_c\to\infty\) the model behaves as a CFT for a majority of its correlation functions, except for those  having the shortest initial or intermediate states of length \(L=2 \), which correspond to the operators \(\Tr(\phi_i\phi_j)\) or  \(\Tr(\phi_i\phi_j^\dagger)\). 
 These states induce  singularities related to the double-trace couplings of the terms \(\Tr(\phi_i\phi_j)\Tr(\phi_i^\dagger\phi_j^\dagger)\) and \(\Tr(\phi_i^\dagger\phi_j)\Tr(\phi_i\phi_j^\dagger)\), which are generated in the action by the renormalization group. Even in the planar limit, the double-trace couplings are running with the renormalization scale \cite{Fokken2014a,Fokken2014b,Fokken2014c,Sieg:2016vap}. However, the coupling \(\xi\) does not run in the planar limit, which preserves the conformal behavior of the majority of physical quantities of the theory.   
   
     The choice of imaginary parameters \(\gamma_j\) in this theory makes it non-unitary, and the double scaling limit introduces a certain chiral structure (orientation of vertices) on the planar Feynman graphs, suggesting  the name \(\chi\)FT for such a QFT. The non-unitarity is a price to pay for one of the remarkable features of the bi-scalar theory --- a great simplification of its Feynman graph content: for most of the physical quantities, each loop order of planar perturbation theory contains at most one Feynman graph. The bulk part of a sufficiently large graph is always of  ``fishnet'' type --- it  consists of a large chunk of regular square lattice. The vertical and horizontal   lines of this lattice  correspond to the lines of propagators of the fields \(\phi_1\) and \(\phi_2\), respectively. Hence, this bi-scalar \(\chi\)FT\(_4\) represents a field-theoretical realization of  A.~Zamolodchikov's integrable statistical mechanics model of fishnet graphs~\cite{Zamolodchikov:1980mb}.       The presence of two flavors leads to a rich set of possibly integrable, i.e.\ potentially computable, individual 4D scalar Feynman graphs \cite{Gurdogan:2015csr,Caetano:2016ydc}.

The purpose of this paper is to study the planar scalar amplitudes in the theory defined by \eqref{Lthree},  and to establish their  explicit Yangian symmetry at any loop order. Since these amplitudes appear to be free of any divergencies, UV or IR, the statement of Yangian symmetry will be precise and, potentially, exploitable for computations of particular Feynman graphs of fishnet type with disc topology.     

 The most general quantities  which  we will study here,  and for which we will establish the Yangian symmetry, are  the following single-trace correlators  \begin{align}\label{TrCorrelator}
K(x_1,x_2,\dots x_{2M})=
\left<\Tr\left[\chi_{1}(x_1)\,\chi_{2}(x_2),\dots \chi_{2M}(x_{2M})\right]\right>,
\end{align}   where \(\chi_{i}\in\{\,\phi_1^\dagger ,\phi_2^\dagger ,\phi_1,\phi_2\}\). In momentum space, this correlator can be interpreted as a scattering amplitude of very massive scalar ``Higgs'' particles which interact with each other only due to the exchange of  massless scalars \(\phi_1,\phi_2\), with momenta much smaller than the Higgs masses.

\begin{figure} 
\begin{center}
\includegraphics[width = 0.8 \textwidth]{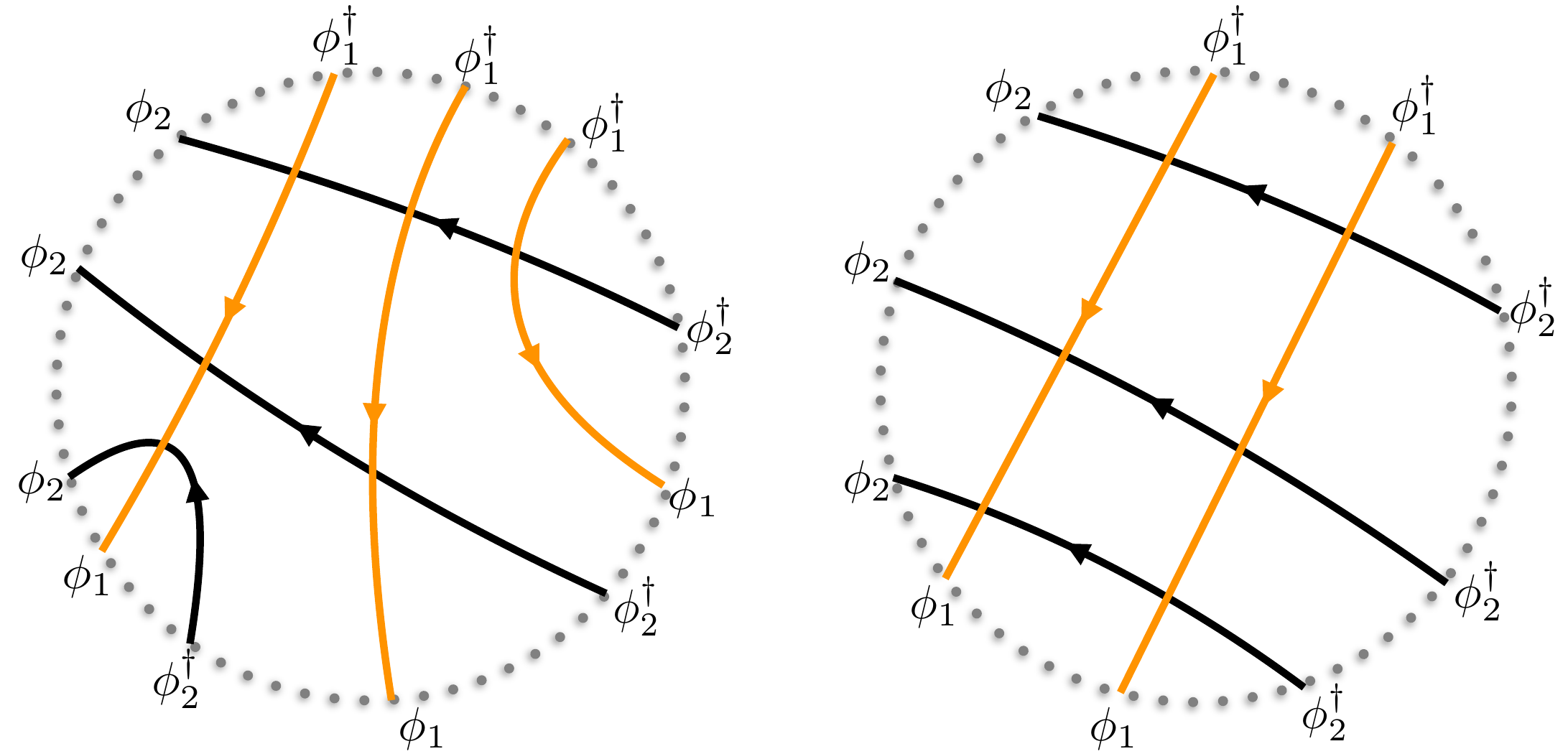}
\end{center}
\caption{\label{Fig:biscalar} Bi-scalar amplitude diagrams. (A): $M_1 = M_2 = 3$ (B): $M_1 =2, M_2 = 3$. }
\end{figure}

\begin{figure}
\begin{center}
\includegraphics[scale=1]{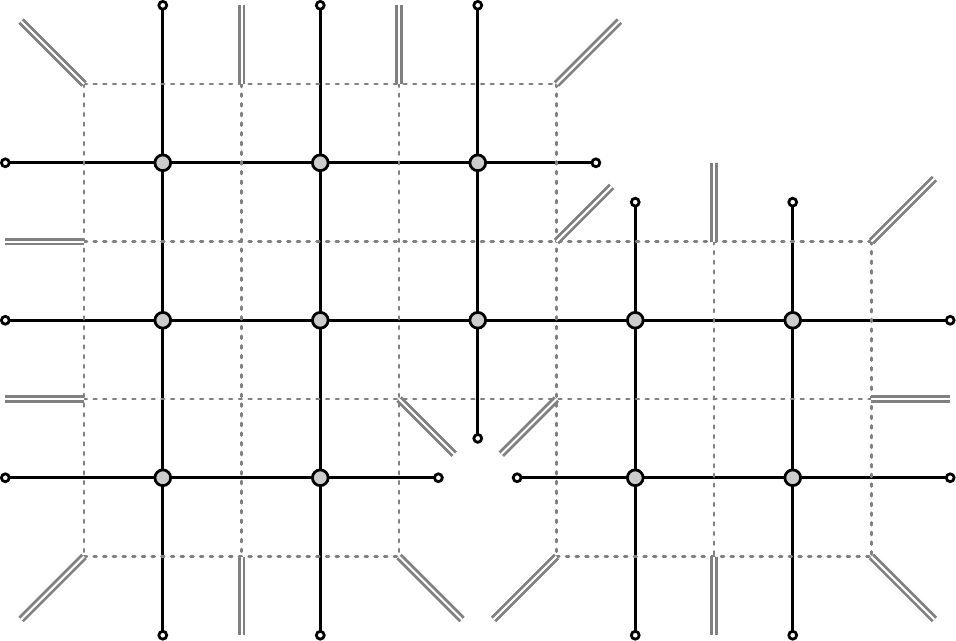}
\end{center}
\caption{ \label{Fig2}
A generic fishnet graph with regular boundary. It is drawn by solid lines. It depends on a number of variables $x_i^{\mu}$ which are coordinates of external legs. Each solid line of the fishnet graph represents a scalar propagator $x_{ij}^{-2}$. Integrations are over positions of vertices (denoted by filled blobs). The dual graph is drawn by dotted lines. The dual graph lives in the momentum representation with integrations over loop momenta. Its external momentum variables are defined as $p^\mu_i = x^\mu_i -x^\mu_{i+1}$. The dual graph does not necessarily correspond to an amplitude in the bi-scalar theory, since it could have interaction vertices of valency
different from four. The external legs of the dual graph are amputated. The inflowing off-shell momenta are denoted by double lines.
}
\end{figure}

Obviously, due to charge conservation, the number of fields \(\phi_1\) (which we denote by \(M_1)\) should be equal to the number of fields \(\phi_1^\dagger\), and the number \(M_2\) of fields \(\phi_2\) should be equal to the number of fields \(\phi_2^\dagger\), so that \(M=M_1+M_2\).   The interaction vertex conserves each of the two flavors. Hence, in the corresponding Feynman graph of disc topology, the propagator lines of  fields \(\phi_1\)  continuously go from one external leg to another, and similarly for \(\phi_2\). 
Lines of the same flavor never cross and the intersection of two types of lines can only happen with one orientation: say the arrows on propagators of fields \(\phi_1\) and \(\phi_2\) around any vertex should follow in the clock-wise order,  as is shown in Fig.~\ref{Fig:biscalar}.
 It is easy to convince oneself that the only possible planar graphs of this type correspond to a disc cut out of the regular square lattice  along a sequence of \(2M_1+2M_2\) edges, as depicted by solid lines in Fig.~\ref{Fig2}.\footnote{This fact was mentioned in the conclusions of \cite{Caetano:2016ydc}.} 
 
\begin{figure} 
\begin{center}
\includegraphics[width = 0.8 \textwidth]{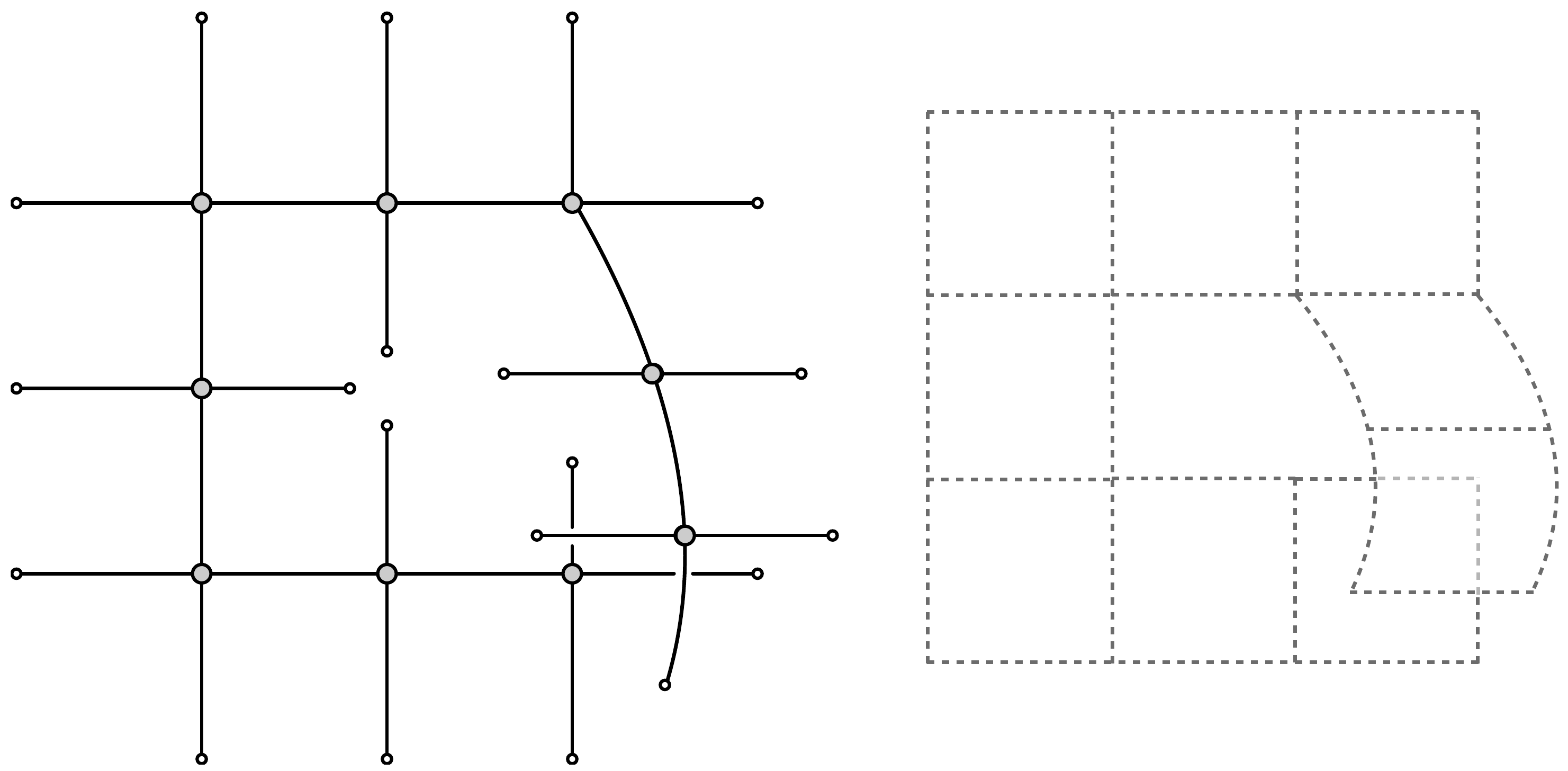}
\end{center}
\caption{\label{FigOverlap} 
Correlator diagram and its dual (without external legs), which cannot be cut out of a simple sheet of regular square lattice. However, it can be cut out of a ``double-sheet'' regular square lattice with a branchpoint. More general graphs can be cut out of the lattices having various ``conical'' singularities, see footnote~\ref{ftnt3}.}
\end{figure}

 The two types of  lines   are made out of propagators of the field \(\phi_1  \) and the  field \(\phi_2\), respectively. 
Each such graph corresponds to a certain ordering of legs around the boundary, chosen from the set \(\{\phi_1^\dagger ,\phi_2^\dagger ,\phi_1,\phi_2\}\). For a given ordering, there is a single possible graph. This means that each single-trace correlator of scalar fields in the planar approximation of the bi-scalar  \(\chi\)FT\(_4\)    is described by a single graph, whose loop order equals the number of interaction vertices inside the disc.
 The number of intersections in turn, depends only on $M_1, M_2$ and on the ordering of the scalar fields under the trace.%
 \footnote{\label{ftnt3} Note that such a diagram of disc topology, although made out of the regular square lattice, can not always be drawn on the plane without overlaps, as can be seen in Fig.~\ref{FigOverlap}. The most general diagram can  be cut out of a  square lattice having various conical singularities, such as branchpoints, in analogy with the Riemann surface.  We will call them  {\it singular square lattices}. The irregularities should appear only at the boundary of this disc, the bulk being always a regular square lattice. } 
In the following, we will mostly drop the arrows on propagator lines since our symmetry considerations apply to generic Feynman graphs of fishnet type --- independently of their origin in a particular theory. Adding the arrows, however, is useful in order to illustrate that the \(\chi\)FT\(_4\) furnishes a generating theory for all of these diagrams, with a one-to-one correspondence between correlators and Feynman graphs.

We will show in the next section of this paper
that the correlator  \eqref{TrCorrelator}, and hence its only Feynman diagram, obeys a Yangian symmetry.

\begin{figure} 
\begin{center}
\includegraphicsbox[scale=1]{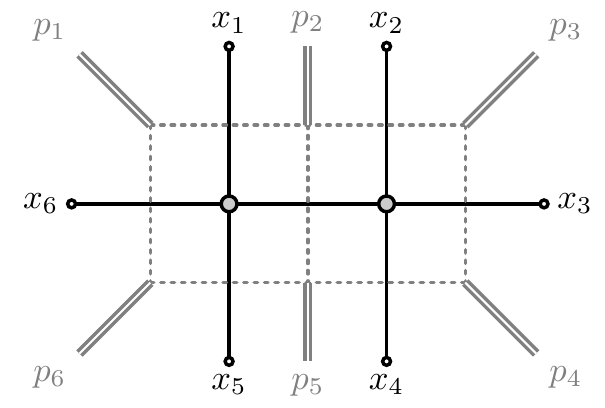}
\hfill
\includegraphicsbox[scale=1]{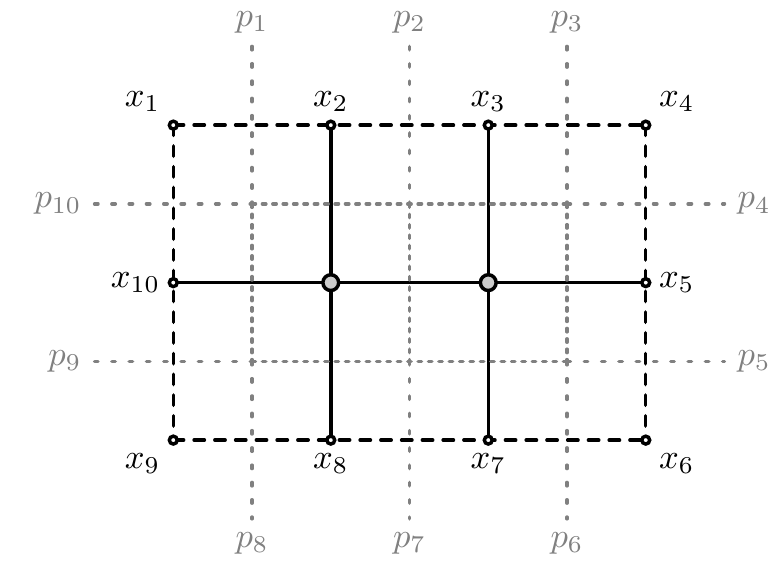}
\end{center}
\caption{\label{Fig1} Two-loop diagrams: double box topology in momentum variables (integration over momenta flowing in loops) and its dual double cross topology in region momentum variables (integration over position of the vertices -- filled blobs). (A): All inflowing momenta (depicted by double lines) are off shell, $p^2_{i} = x^2_{i\,i+1} \neq 0$ at $i = 1, \ldots, 6$. (B): Amplitude diagram for scattering of massless particles, i.e.\ the inflowing momenta (depicted by loosely dotted gray lines) are light-like $p^2_{i} = x^2_{i\,i+1} =0$ at $i = 1, \ldots, 10$. These constraints are imposed by means of delta functions $\delta(x_{i \, i+1}^2)$ depicted by dashed black lines.}
\end{figure}

The  correlator  \eqref{TrCorrelator}   is our master object from which we can obtain any amplitude. It is more general than just the color-ordered scattering amplitudes of massless bosons, but the latter can always be recovered from these correlators by passing to dual momentum variables  \(p_i=x_{i}-x_{i+1}\) and by putting all \(p_i\)'s on the light cone, as shown on the particular example of the double-cross diagram in the right of 
Fig.~\ref{Fig1}. This duality transformation  can be  represented graphically as a passage to the dual lattice, depicted by dotted lines on  the right of  Fig.~\ref{Fig1}. It  represents a possible bi-scalar amplitude with quartic interaction vertices.
The internal  momenta to be integrated, correspond to square faces.  
    To put  the \(i\)th leg of this dual graph on shell, we simply cut the corresponding propagator by replacing it in coordinate space via  $x_{i\,i+1}^{-2} \to \delta(x_{i\, i+1}^2)$ or in momentum space according to $p_{i}^{-2} \to \delta(p_{i}^2)$. Hence, we  amputate this external propagator as prescribed by the Lehmann--Symanzik--Zimmermann (LSZ) procedure.

\begin{figure}
\begin{center}
\includegraphics[scale=1]{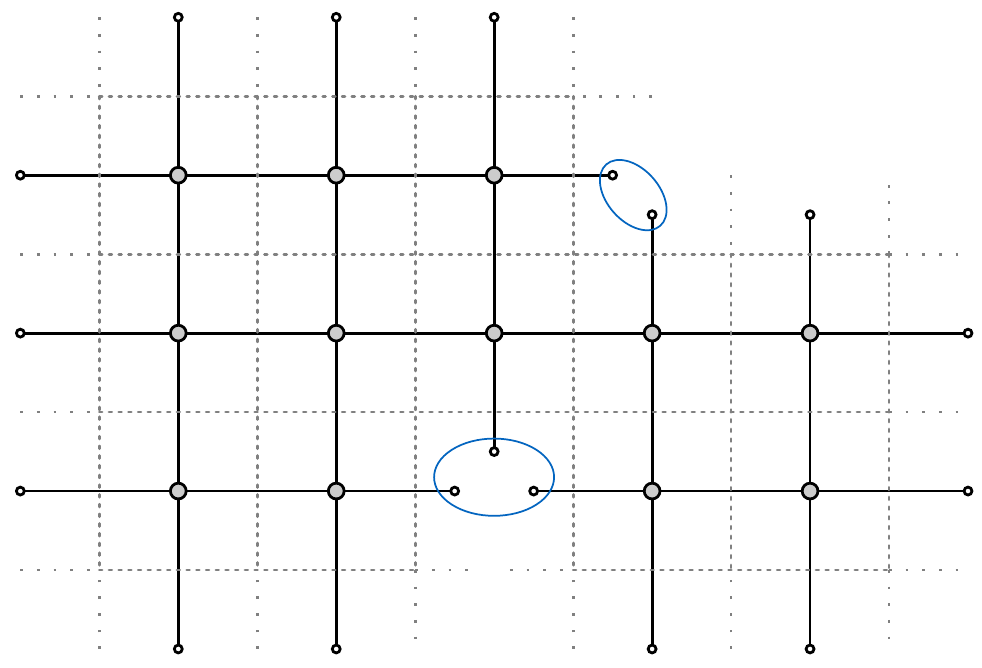}
\end{center}
\caption{ \label{Fig2-2} In this picture, a bi-scalar scattering amplitude of massless scalars is represented in the dual graph depicted by dotted lines. It has doubled (w.r.t.\ the previous Fig.~\ref{Fig2}) external legs at the convex corners of the boundary, and no legs at the concave corners. The corresponding additional momentum variables at convex corners should satisfy the momentum conservation condition at each dual vertex.  We obtain the admissible bi-scalar amplitude  by identifying  the   points of the boundary of the original graph which end in the same square (i.e., the same site of the original lattice. Such points are surrounded by  ellipses in the above figure. The masslessness is ensured by additional factors \(\delta(p_j^2)\) multiplying the external legs of the dual graph.
}
\end{figure}

Notice that in the left Fig.~\ref{Fig1} the dual graph with double-line legs represents a particular double-box amplitude,  where the external vertices can have three neighbors (and the external legs can be on shell or not). A more general dual  graph in Fig.~\ref{Fig2} can also have five neighboring propagators at some external vertices.  Such  graphs do not represent amplitudes in the bi-scalar theory defined by
\eqref{Lthree}
since the number of particles of each of the two flavors cannot be conserved.
However, we can recover from the graph in Fig.~\ref{Fig2} a  bi-scalar amplitude by doubling the external momentum legs at the convex corners and omitting them at concave corners of the dual graph, which corresponds to joining some external legs,   as shown on the Fig.~\ref{Fig2-2}. Such a graph, with some external coordinates identified, will be called ``irregular'' in the following.

Importantly, it can be argued that all of the above fishnet-type amplitudes are actually finite, both in the IR and UV regime. 
This means that, unlike  the case of loop amplitudes in \(\CN=4\) SYM theory, the conformal and the full Yangian symmetry of these amplitudes can be taken at face value.  The rigorous establishment of this Yangian invariance, which is probably a close relative of the elusive (twisted) Yangian  invariance of the full \(\gamma\)-deformed  \(\CN=4\) SYM theory, is the main result of this paper.  First, we will demonstrate this invariance by application of a specific monodromy matrix to the external legs around the Feynman graphs under consideration (the ``lasso'' method). Then, expanding it with respect to the spectral parameter, we obtain the level-one generators of the conformal Yangian algebra and comment on the relation to dual conformal symmetry.

\section{Yangian Invariants and Monodromy Matrix}

In this section we briefly review the construction of Yangian invariants before specifying it to the case of the conformal algebra $\alg{so}(2,4)$.
Historically, the so-called RTT realization of the Yangian algebra first appeared implicitly in the context of the quantum inverse scattering method \cite{Faddeev:1996iy,Sklyanin:1991ss}, much earlier than its general definition by Drinfel'd. In this framework, the Yangian commutation relations are encoded into a Yang--Baxter equation for the monodromy matrix $T(u)$, the so-called RTT-relation:
\begin{align} \label{RTT}
R(u-v) \; T(u)\otimes T(v) = T(v)\otimes T(u) \;R(u-v)\,.
\end{align} 
The monodromy matrix $T(u)$ is a formal series in the spectral parameter $u$. 
It encompasses the infinite set of Yangian generators $\Jexp^{(n)}_{\alpha\beta}$ \cite{Molev:1994rs}:
\begin{align} \label{TJ}
T_{\alpha\beta}(u) = \delta_{\alpha\beta} + \sum^\infty_{n = 0} u^{-n-1} \Jexp^{(n)}_{\alpha\beta}\,.
\end{align} 
In the case of our interest, $\alpha,\beta =1 ,\ldots ,4$ are matrix indices of the defining representation of $\alg{su}(2,2) \approx \alg{so}(2,4)$,
and the numerical matrix $R(u)$ is Yang's R-matrix
\begin{equation}
R(u) = {\bf 1} + u \, P\,,
\end{equation} 
where $P$ is the permutation matrix. It satisfies the Yang-Baxter equation which is a consistency relation 
for the structure constants of the Yangian algebra given by the RTT-relation.

The evaluation representation with $ \Jexp^{(n)} =0$ for $n > 0$  provides the simplest solution of (\ref{RTT}), namely
$T_{\alpha\beta}(u) = \delta_{\alpha\beta} + u^{-1} \Jexp^{(0)}_{\alpha\beta}$. For the case at hand, the
$ \Jexp^{(0)}_{\alpha\beta}$ are linear combinations of the conformal algebra generators acting on a single site of a non-compact spin chain. 
We call the above solution the \emph{Lax operator} and denote it by $L(u)\equiv \delta_{\alpha\beta} + u^{-1} \Jexp^{(0)}_{\alpha\beta}$.%

 The algebra given by the RTT-relation possesses a comultiplication structure.
In particular, this means that the matrix product of several Lax operators (each acting on its own spin chain site) respects (\ref{RTT}).
We thus find that the inhomogeneous monodromy $T(\vec u) = L_n(u_n) \ldots L_2(u_2) L_1(u_1)$ specified by the set of 
parameters $\vec u = (u_n, \ldots, u_1)$ furnishes a solution of eq.\ (\ref{RTT}). Here $L_i(u_i)$ acts on the $i$-th site of the spin chain.
In the following we will change the normalization of the $n$-site monodromy (\(L_j(u)\to uL_j(u)\)) in order to make it polynomial in the spectral parameter.
 
As was shown in \cite{Chicherin:2013sqa,Chicherin:2013ora}, the eigenvalue problem for an inhomogeneous monodromy
constructed out of Lax operators, i.e.\
\begin{align} \label{LLL}
L_n(u_n) \ldots L_2(u_2) L_1(u_1)\, \ket{\lambda;\text{inv}} = \lambda(\vec u) \,  \ket{\lambda;\text{inv}} \idop,
\end{align}
provides a natural way to obtain Yangian invariants $\ket{\lambda;\text{inv}}$, which live on $n$ sites of a non-compact spin chain. 
Both sides of eq.\ (\ref{LLL}) are matrices and $\idop$ denotes the identity matrix.
Here $u_i = u +\delta_i$ is the spectral parameter $u$ shifted by the inhomogeneity $\delta_i$.
The invariants are functions of 
$({\bf a}_1,\ldots,{\bf a}_n)$,
where ${\bf a}_i$ is a set of spin variables on the $i$-th site, and they depend on the inhomogeneities $\delta_i$. 
$\lambda(\vec u)$ is a polynomial in $u$ of degree $n$.
Eq.\ (\ref{LLL}) implies that the non-diagonal Yangian generators, obtained from the expansion (\ref{TJ}),
annihilate $\ket{\lambda;\text{inv}}$ and that the diagonal generators act covariantly on it, i.e.\
\begin{align*}
\Jexp^{(n)}_{\alpha\beta} \ket{\lambda;\text{inv}}  =  c_{n}\delta_{\alpha\beta}\, \ket{\lambda;\text{inv}} 
\end{align*}
with some coefficients $c_{n} = c_n(\vec\delta)$ specified by the expansion of the polynomial $\lambda(\vec u)$ in powers of $u$.
In \cite{Chicherin:2013sqa} the eigenvalue problem (\ref{LLL}) for a non-compact spin chain with 
Jordan--Schwinger representations of the algebra $\alg{sl}(N)$ was studied, and the R-operator method for the superconformal algebra $\alg{gl}(4|4)$ was investigated in \cite{Chicherin:2013ora,Frassek:2013xza,Kanning:2014maa,Broedel:2014pia,Kirschner:2014pza} to construct tree-level scattering amplitudes in ${\cal N}=4$ SYM, as well as in ABJM theory \cite{Bargheer:2014mxa}. 
The R-operator method of \cite{Chicherin:2013ora} was also applied to construct form factors of composite operators \cite{Bork:2015fla,Frassek:2015rka}, the form factors of Wilson lines \cite{Bork:2016xfn}, the kernels of QCD parton evolutions \cite{Fuksa:2016tpa}, amplituhedron volume functions \cite{Ferro:2016zmx}, and splitting amplitudes \cite{Kirschner:2017vqm}.

Here we apply the eigenvalue problem (\ref{LLL}) to a different physical setup. We consider the principal series representation of the conformal algebra $\alg{so}(2,4)$ of four-dimensional Minkowski space. We are going to show that any fishnet graph $G$ representing the correlator (\ref{TrCorrelator}) solves the relation (\ref{LLL}) for an appropriate choice of parameters $u_i = u + \delta_i$, i.e.\ it is an eigenstate (which we denote from now on by $\ket{G}$) of the $n$-site monodromy. The role of the spin chain sites is played by $n$ external legs (or vertices) of a scalar amplitude, and the spin chain variables $x_i^{\mu}$ represent the 4D coordinates at each site. More precisely, the $x_i^{\mu}$ are region momenta for planar amplitudes. For an $n$-particle amplitude they are defined as usual via $p^\mu_{i} = x_i^\mu - x_{i+1}^\mu$, where $i = 1, \ldots,n$ and $x_{n+1} \equiv x_1$. So they have the same dimension as the momenta. We can also think of the fishnet graphs as of correlator diagrams. In this case $x_i^{\mu}$ are the usual position space coordinates.

\section{Conformal Lax Operator}
\label{sec:conf_Lax}

In order to build the Yangian generators, we need to specify the Lax operator for the conformal algebra.
Considering scattering amplitudes with massless legs we have to stick to Minkowski signature, while for amplitudes with all external legs massive or for correlation functions we might also use Euclidean signature. To be specific, let us choose Minkowski signature, so that
the conformal algebra is $\alg{so}(2,4)$. Let us denote the generators of $\alg{so}(2,4)$ by $M_{ab}$, $a,b=1,\ldots,6$.

We will need the differential representation $\rho_{\Delta}$ of the conformal algebra generators on the space of scalar fields carrying conformal dimension $\Delta$. 
In this representation the generators 
\begin{equation}
\rho_{ab} \equiv \rho_{\Delta}(M_{ab})
\end{equation}
take the familiar form.
They are given by the following first order differential operators of translations $(P_{\mu})$, dilatations $(D)$, Lorentz rotations $(L_{\mu\nu})$, and conformal boosts $(K_{\mu})$:
\begin{align}
&D = -i x_{\mu} \partial_{x_{\mu}} - i \Delta \;,\;\;\;
L_{\mu\nu} = i x_{\mu}  \partial_{x^{\nu}}-i x_{\nu}  \partial_{x^{\mu}} \;,\; \nonumber \\
&P_{\mu} = - i \partial_{x^{\mu}} \;,\;\;\;
K_{\mu} = 2 \, x^{\nu} \, (L_{\nu \mu}) - i (x^{\nu} x_{\nu})  \partial_{x^{\mu}} - 2 i \Delta x_{\mu}.
\label{eqn:diffrepconfgen}
\end{align}
This is a particular case of the principal series representation of weight \((\Delta,0,0)\). The generators depend on one 
representation label $\Delta$ which  is the conformal dimension (or conformal spin).

We will also need an irreducible spinor representation of $\alg{so}(2,4)$. 
To construct these generators we use gamma-matrices $\Gamma_a$ for the six-dimensional space $\mathbb{R}^{2,4}$. 
These are matrices in the eight-dimensional space $V$. Then we consider their commutators $\frac{i}{4}[\Gamma_a,\Gamma_b]$, which
provide a spinor representation of $\alg{so}(2,4)$. However, the latter take block diagonal form, i.e.\ they form a reducible representation 
with $V = V_{+} \oplus V_-$. Taking the Weyl projection onto $V_+$ we obtain the irreducible four-dimensional representation 
\begin{equation}
s_{ab} \equiv s(M_{ab}) = \frac{i}{4}[\Gamma_a,\Gamma_b]\Big|_{\downarrow V_+}. 
\end{equation}
It is equivalent to the defining representation of $\alg{su}(2,2) \approx \alg{so}(2,4)$. 

Now we have all necessary ingredients to write down the main operator for all our considerations.
We use the Lax operator for the conformal algebra in the form \cite{Chicherin2013}
\begin{align}\label{eq:Lax1}
L_{\alpha\beta}(u;\Delta) = u \,\delta_{\alpha\beta}+ \frac12 s^{ab}_{\alpha\beta}\,\rho_{ab}\,,
\end{align}
which is a $4\times 4$ matrix ($\alpha,\beta$ are matrix indices) with first order differential operator entries.
It has a nice factorized form \cite{Chicherin2013}:
\begin{align} \label{Lax}
L( u_+,u_- ) = \left(
\begin{array}{cc}
  \boldsymbol{1} & \; \mathbf{0} \\
  \mathbf{x} & \; \boldsymbol{1}
\end{array}
\right) \left(
\begin{array}{cc}
  u_+ \cdot \boldsymbol{1}  & \ \ \mathbf{p} \\
  \mathbf{0}  & u_- \cdot \boldsymbol{1}
\end{array}
\right) \left(
\begin{array}{cc}
  \boldsymbol{1} & \; \mathbf{0} \\
  - \mathbf{x} & \; \boldsymbol{1}
\end{array}
\right) ,
\end{align}
where the block $2 \times 2$ matrices are ${\bf x} \equiv - i \overline{ \boldsymbol{\sigma} }_{\mu} x^{\mu}$ and  ${\bf p} 
\equiv -\frac{i}{2}\boldsymbol{\sigma}_{\mu} \partial_{x_{\mu}}$.
The numbers
\begin{align} \label{u+u-}
u_+ \equiv u + \frac{\Delta - 4}{2} \qquad, \qquad u_- \equiv u - \frac{\Delta}{2}
\end{align} 
are linear combinations of the spectral parameter and the conformal dimension. Let us note that $\Delta = u_+ - u_-  +2$.
In the following we will use both notations, $L(u_+,u_-)$ and $L(u;\Delta)$, for the Lax operator (\ref{Lax}).

In Minkowski signature we have
$\boldsymbol{\sigma}_{\mu} = ( \boldsymbol{1} ,  \sigma_1, \sigma_2, \sigma_3 )$ and 
$\overline{ \boldsymbol{\sigma} }_{\mu} = ( \boldsymbol{1} , - \sigma_1, - \sigma_2, - \sigma_3 )$, 
where $\sigma_{i}$, $i=1,2,3$ are the standard Pauli matrices. 
The Lax operator (\ref{Lax}) satisfies the RTT-relation (\ref{RTT}). 
For more details about the conformal Lax operator see \cite{Chicherin2013}.

Let us abbreviate $x_{ij} \equiv x_i - x_j$, and $x^2 \equiv x_{\mu} x^{\mu}$ with Minkowski signature.
We will need the following properties of the Lax operator (\ref{Lax}):
\begin{itemize}
\item 
We denote by $L^T$ the transposed Lax operator in the non-compact physical space (not the auxiliary matrix space), i.e.\ we have
$x_{\mu}^T = x_{\mu}$ and $\pa_{x_{\mu}}^T = - \pa_{x_{\mu}}$, which is equivalent to integration by parts.  
The inverse of the Lax operator coincides with its transposition (up to permutation and shift of parameters)
\begin{align} \label{transpose}
L^{T}_{\alpha\beta}(v-2,u-2) L_{\beta\gamma}(u,v) = u v \,\delta_{\alpha\gamma}\,.
\end{align}
\item The Lax operator acts diagonally onto $1$ at $\Delta = 0$:
\begin{align} \label{vacuum}
L_{\alpha\beta}(u,u+2) \cdot 1 = (u+2) \delta_{\alpha\beta}\;,\;\;\; L^T_{\alpha\beta}(u+2,u) \cdot 1 = (u+2) \delta_{\alpha\beta}\,.
\end{align}
We can think of $1$ as of a local pseudo-vacuum state of the Lax operator. 
Starting from this vacuum, we then construct  non-trivial states of the non-compact spin chain.
\item
The scalar propagator $x_{12}^{-2}$ is an intertwining operator permuting
spectral parameters of the two-site monodromy \cite{Chicherin2013,Derkachov2006b,Derkachov2009,Derkachov:2010zz}:
\begin{align} \label{intw}
x_{12}^{-2} L_1(u,v) L_2(w,u+1) = L_1(u+1,v) L_2(w,u) x_{12}^{-2}.
\end{align}
\item Working in Minkowski signature, we 
can consider $\delta(x_{12}^2)$,  which is the unitary cut of the Feynman propagator $1/(x^2_{12} + i \epsilon)$. 
It satisfies the same intertwining relation:
\begin{align}
\delta(x_{12}^2) L_1(u,v) L_2(w,u+1) = L_1(u+1,v) L_2(w,u) \delta(x_{12}^2).  \label{intwdelt}
\end{align}
\item 
The eigenvalue problems for monodromies with the same cyclic ordering of Lax operators are equivalent  (cyclicity) \cite{Chicherin:2013sqa}:\begin{align}
L_n(u_{n};\Delta_{n}) \ldots L_1(u_{1};\Delta_{1})\, \ket{G} &= \lambda \,  \ket{G} \idop  \; \notag \\
 &\Updownarrow\notag\\
L_{n-1}(u_{n-1};\Delta_{n-1}) \ldots L_1(u_{1};\Delta_{1}) L_n(u_{n}-4;\Delta_{n})\, \ket{G} 
&= \widetilde\lambda\,  \ket{G}  \idop,\label{cyclT}
\end{align}  where by \(u_k\) we mean different spectral \(u\)-parameters for each Lax operator.%
\footnote{Their relation to the below overall spectral parameter $u$ of the monodromy (see e.g.\ the $u$ in \eqref{eq:notation}) is given by $u_k=u+a_k$ with some shift $a_k$.}
Here the eigenvalues $\lambda$ and $\widetilde\lambda$ are related by $u_{n+} u_{n-} \widetilde\lambda = (u_{n+}-2) (u_{n-}-2) \lambda$. 
For the benefit of the reader, in Appendix \ref{AppCycl} we adapt the proof from \cite{Chicherin:2013sqa} to the conformal Lax operator (\ref{Lax}).
Implications of cyclicity for the first realization of the Yangian generators are discussed in Section \ref{sec:firstreal}.
\end{itemize}

\section{Yangian Symmetry of Bi-Scalar Fishnet Graphs: Regular Boundary}
\label{sec:YangianSymmetry}

\begin{figure} 
\begin{center}
\includegraphics[scale=1]{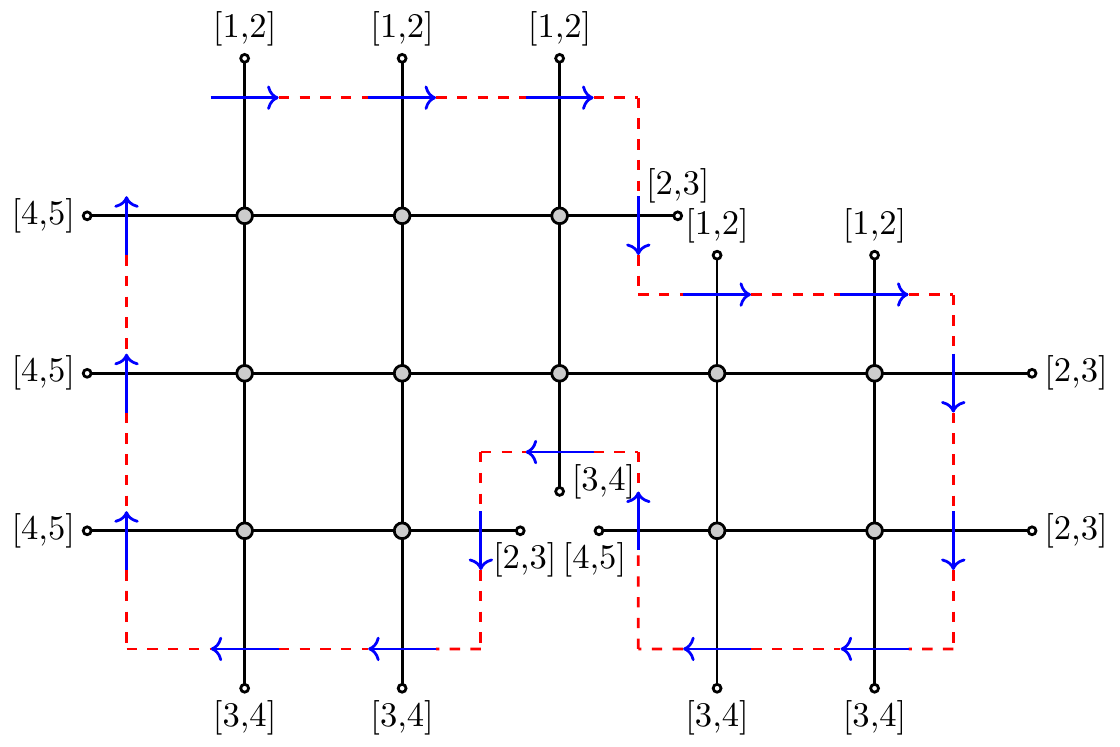}
\end{center}
\caption{\label{Fig3} We act with the monodromy, i.e.\ with the ordered matrix product of conformal Lax operators on the left hand side of eq.\ (\ref{main}), 
on the fishnet graph from Fig.~\ref{Fig2}. 
The monodromy is depicted by the oriented contour. Lax operators correspond to solid blue segments of the contour and dashed red lines denote summation over matrix indices. The contour is decorated by inhomogeneities $[\delta_{i}^+,\delta_i^-]$ of the monodromy, which indicate the shifts in the arguments of Lax operators $L(u+\delta^+_i,u+\delta^-_i)$, cf.\ eq.\ (\ref{Lax}).}
\end{figure}

Now we consider fishnet graphs for the correlators \eqref{TrCorrelator} and we show that they are invariants of the Yangian algebra. 
As it was established above, they have a disc topology with  the square lattice, ``fishnet'' structure in the bulk. 
For simplicity we start with fishnet graphs without junction of external legs, see Fig.~\ref{Fig2}. 
They are correlator graphs with all external points $x_1,\ldots,x_{2M}$ in eq.\ (\ref{TrCorrelator}) chosen to be different. 
We call such a boundary {\it regular}. 
These graphs are simpler to comprehend than the generic, irregular case discussed in the subsequent sections. 
The graphs in Fig.~\ref{Fig:biscalar} have a regular boundary.

We will demonstrate the Yangian symmetry and its proof for an arbitrary planar fishnet graph with $2M$ disjoint external legs using a rather  generic example of such a graph  drawn by solid lines in Fig.~\ref{Fig2}. It lives in the coordinate representation. As we explained in the introduction, it can be cut out of a rectangular lattice by means of scissors, cutting a sequence of edges along a closed line forming the boundary. 
This boundary is simply-connected  (disc topology, no holes inside the graph) and it forms a polygon with all angles right. 
The coordinates of the external legs are arbitrary, so that the graph represents a multi-loop multi-variable Feynman integral. Four-dimensional integrations are assigned to quartic vertices (denoted by filled blobs in Fig.~\ref{Fig2}). The integral depends on variables $x^\mu_i$, $i=1,\ldots,2M$, which are coordinates of external legs. These external points of the diagrams are depicted by small white blobs. We can think of the variables $x^\mu_i$ in two ways. We can consider them as true $x$-space coordinates, and in this case the graph is a correlator diagram corresponding to eq.\ (\ref{TrCorrelator}). 

We can also think of the variables $x^\mu_i$ as of region momenta. In this case they are kinematic variables of the dual graph (drawn by dotted lines in Fig.~\ref{Fig2}) which is a momentum-space Feynman integral. Region momenta are attributed to external faces of the graph, as shown on the Fig.~\ref{Fig1}.
Since they are related to the usual momenta $p_i^\mu$ by the duality transformation $x_{i}^\mu - x_{i+1}^\mu = p^\mu_i$  they automatically resolve the momentum conservation constraint, $p^\mu_1 + \ldots + p^\mu_{2M} = 0$. Integrations in the dual graph are attributed to loops. Since all variables $x^\mu_i$ are independent, all momenta $p^\mu_i$ flowing into the dual graph (denoted by solid double lines in Fig.~\ref{Fig2}) are off shell. As  was  noted before, the dual graph (of a fishnet graph with regular boundary) does not necessarily correspond to an amplitude in the bi-scalar theory: all inflowing momenta are off shell and vertices of valency different from four could appear. Nevertheless fishnet graphs with regular boundary describe some amplitudes, for example the $x$-space double cross integral is dual to the $p$-space double-box integral with six off-shell legs, see Fig.~\ref{Fig1}. To embrace the whole set of amplitudes of the bi-scalar theory we will have to consider fishnet graphs with more generic boundary (see Section~\ref{irreg}).

The loop integrals that we consider are finite. By power counting one can see that UV divergences are absent. Since all inflowing momenta are off shell, IR or collinear divergences do not appear.

For the moment we consider a fishnet graph with regular boundary. Then we draw the contour $\cal C$, oriented clockwise along the boundary such that it crosses all $2M$ external legs, see Fig.~\ref{Fig3}. It denotes the monodromy matrix $L_{2M} \ldots L_1$, and the blue segments depict conformal Lax operators (cf.\ eq.\ (\ref{Lax})) forming it. Red segments denote contractions of matrix (auxiliary space) indices. Slightly abusing notations we also denote by $\cal C$ the ordered set of external legs.

We indicate the parameters of the $i$-th Lax operator of eq.\ (\ref{Lax}) using  the shorthand notation 
\begin{equation}\label{eq:notation}
[\delta_{i}^+,\delta_{i}^-]\equiv 
(u_{i+},u_{i-}) \equiv (u + \delta_{i}^{+}, u+\delta_{i}^{-})\,.
\end{equation}
The rule to assign inhomogeneities along the monodromy is the following: 
\begin{itemize}
\item
At the first leg we choose $[\delta_{1}^{+},\delta_{1}^{-}]=[1,2]$ (of course the overall  spectral parameter $u$ is  allowed to be shifted uniformly  in all $u_{i+}$ and $u_{i-}$ along the contour).
\item
We do not change inhomogeneities of Lax operators when moving straight along a horizontal or vertical segment of the contour. 
\item
We increase $\delta_{i+1}^{\pm} = \delta_{i}^{\pm} + 1$   at a convex  corner $i \to i+1$,  when turning by  an angle $\pi/2$.
\item
We decrease $\delta_{i+1}^{\pm} = \delta_{i}^{\pm} - 1$ at a   concave corner $i \to i+1$, when turning by an angle $-\pi/2$. 
\end{itemize}
Let us note that these rules are consistent with the cyclicity in (\ref{cyclT}).
With the above prescription and according to eq.\ (\ref{u+u-}), representations of the conformal algebra carry the same conformal weight $\Delta = 1$ on all sites of the spin chain.
In the following, also the spin chains carrying representations of different conformal weights will naturally arise.

We claim that the fishnet graph $\ket{G}$ is an eigenfunction of the monodromy
\begin{align} \label{main}
\left(\prod_{i \in {\cal C}} L_i[\delta_{i}^{+},\delta_{i}^{-}]  \right) \ket{G} = 
\left(\prod_{i \in {\cal C}_{\text{out}}} [\delta_{i}^{+}][\delta_{i}^{-}]\right)\ket{G}  \idop \,.
\end{align}
The expression for the eigenvalue in the previous formula can be read off from the respective graph. Here we abbreviate 
\begin{equation}
[\delta_k^\pm] \equiv \big(u+\delta_k^\pm\big).
\end{equation}
We split the set of $2M$ external legs into pairs of antipodes. For each pair we apply the following rule: a leg which we encounter first when moving along the monodromy contour (it has a lower number) belongs to the set ${\cal C}_{\text{in}}$, and a leg which we encounter last moving along the contour (it has a higher number) belongs to the set ${\cal C}_{\text{out}}$.
So we decompose the set of external legs according to ${\cal C} = {\cal C}_{\text{in}} \cup {\cal C}_{\text{out}}$.
In the example in Fig.~\ref{Fig3}, we have $2M = 18$ and the eigenvalue in (\ref{main}) equals $\lambda(u)=([3]\times[4])^5([4]\times[5])^4=(u+3)^5(u+4)^{9}(u+5)^4$.

\section{Proof of Yangian Symmetry: The Lasso Method}

\subsection{Cross Integral}

\begin{figure}
\begin{center}
\includegraphics[width = \textwidth]{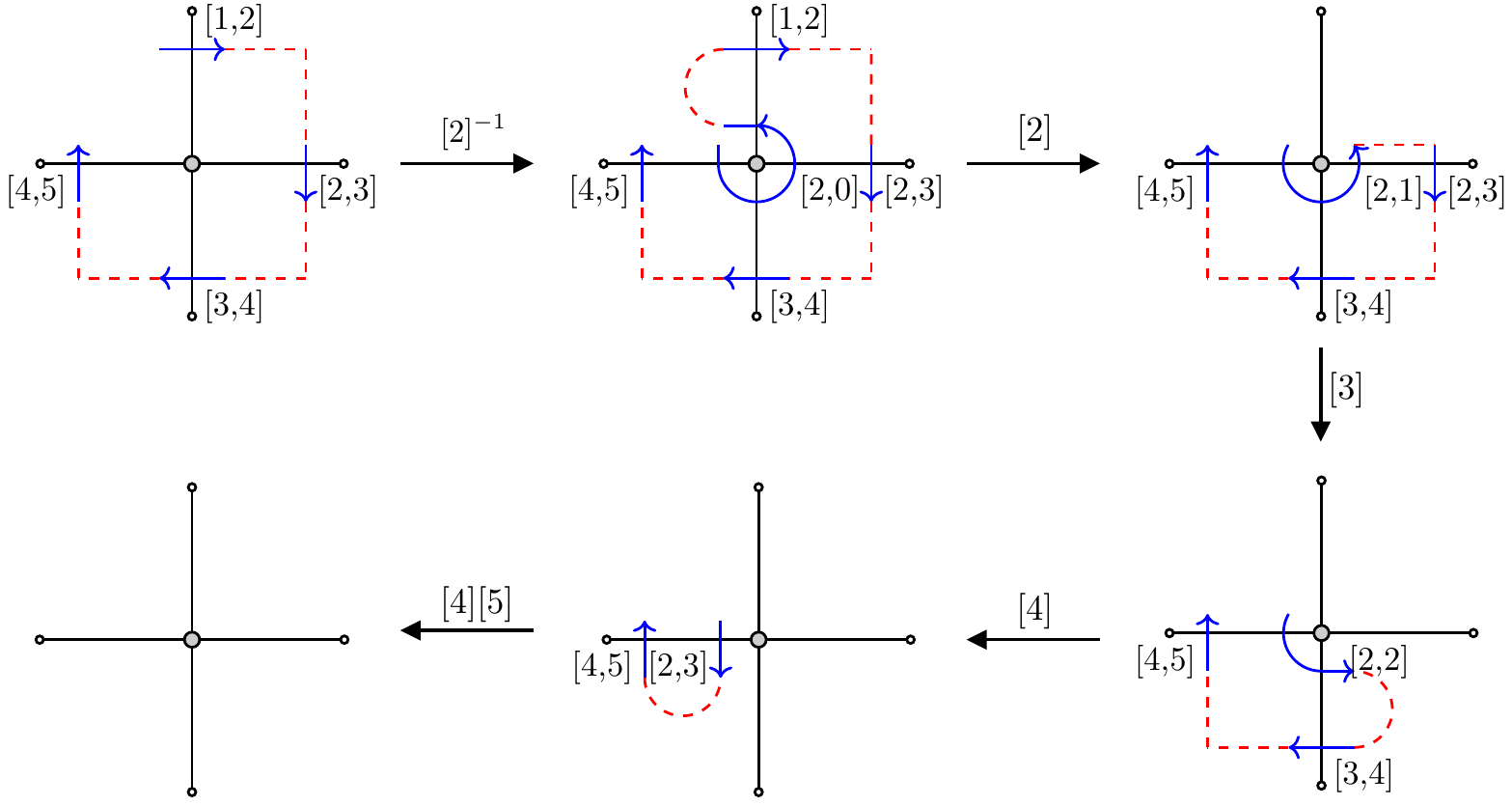}
\end{center}
\caption{\label{crossfig} The sequence of transformations of the monodromy acting onto the cross integral. The  Lax operators of the external part of the monodromy act on fixed external coordinates. The Lax operator introduced around the integrated middle node acts on the coordinate of this node. Numerical factors appearing in the process are indicated above arrows.}
\end{figure}

Before presenting the general proof let us first demonstrate
how it works in several simple cases. We start with the well-known cross integral (one-loop box with four off-shell legs in momentum space):
\begin{align} \label{cross} 
\ket{\text{cross}} = \int \frac{\mathrm{d}^4 x_0}{x_{10}^2 x_{20}^2 x_{30}^2 x_{40}^2}\,.
\end{align}
This integral can be evaluated in terms of dilogarithms \cite{ussyukina1993approach}, and the answer is

\begin{align} \label{DUint-2} 
\ket{\text{cross}} =\frac{\pi^2}{x^2_{13}x^2_{24}} \frac{2\text{Li}_2(z) - 2\text{Li}_2(\bar{z}) + \log z \bar{z} \log \frac{1-z}{1-\bar{z}}}{z - \bar{z}},
\end{align} 
where $z \bar{z} = \frac{x^2_{12} x^2_{34}}{x^2_{13} x^2_{24}}$ and $(1-z)(1-\bar{z}) = \frac{x^2_{14} x^2_{23}}{x^2_{13} x^2_{24}}$.
 We show that the cross integral is an eigenfunction of the monodromy
\begin{align} \label{crossef}
L_4[4,5]L_3[3,4]L_2[2,3]L_1[1,2] \,\ket{\text{cross}} = [3][4]^2[5]\,\ket{\text{cross}}\idop\,.
\end{align}
The proof is given by the series of transformations of the monodromy contour in Fig.~\ref{crossfig}.
Firstly, we extend the ``spin chain'' monodromy by one additional site corresponding to the integration point $x_0$.
More precisely, taking into account eq.\ (\ref{vacuum}), we multiply the four-point monodromy by the Lax operator $L_0^T$ acting on $1$ and integrate it by parts. In this way we obtain the five-point monodromy acting onto the integrand of (\ref{cross}).
Thus, effectively we change the length of the non-compact spin chain and introduce representations with conformal dimension $\Delta$ different from $1$. Then we pull the monodromy through the scalar propagators, which form the integrand, using repeatedly eqs.\ (\ref{vacuum}) and (\ref{intw}).
In Appendix \ref{AppB} we give algebraic expressions corresponding to the sequence of monodromy transformations in Fig.~\ref{crossfig}.

\subsection{Double-Cross Integral}

\begin{figure}
\begin{center}
\includegraphics[scale=1]{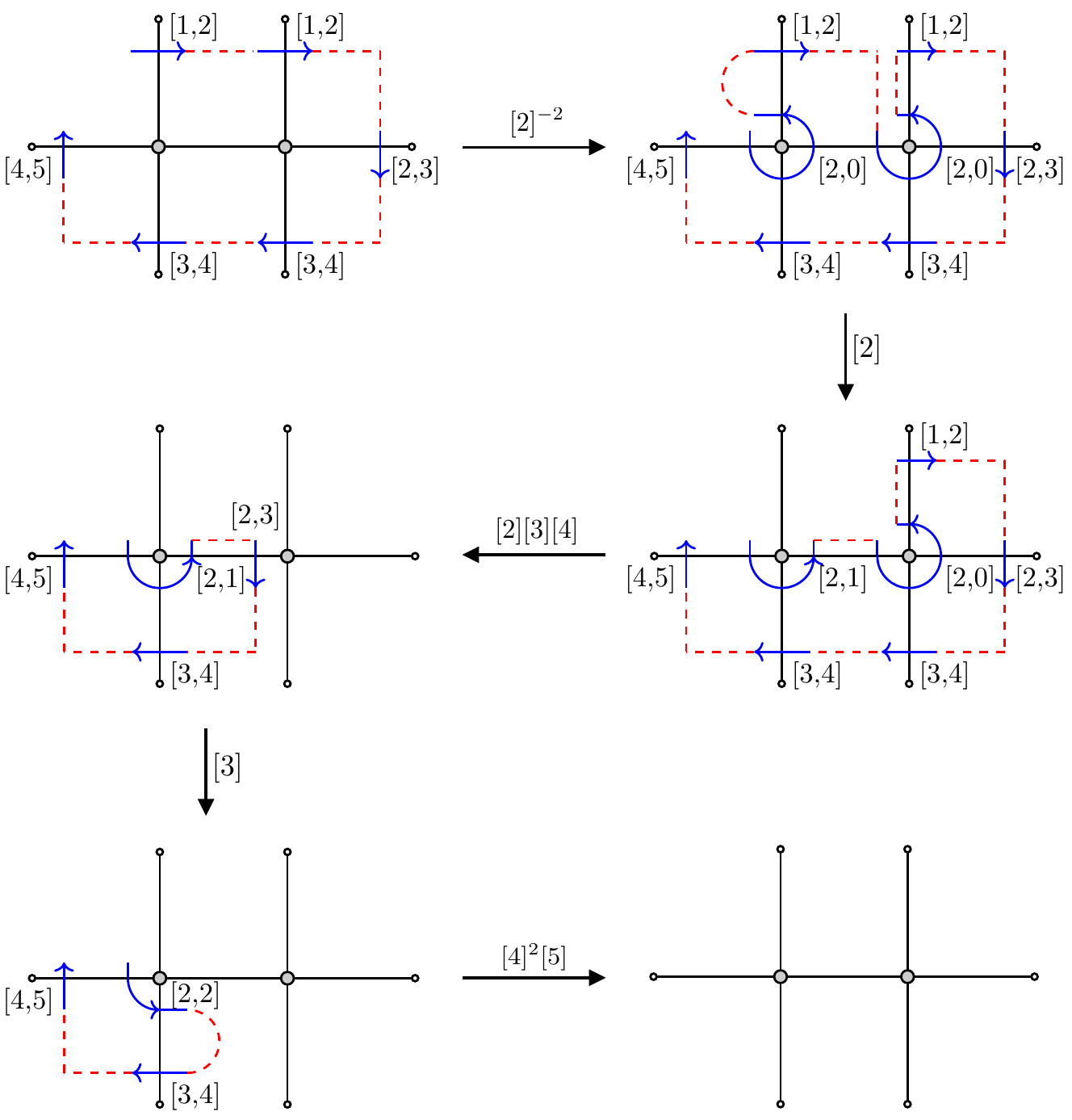}
\end{center}
\caption{\label{figdblcrss} The sequence of transformations of the monodromy acting onto the double cross integral. Numerical factors appearing in the process are indicated above the arrows.}
\end{figure}

Let us consider the double-cross integral (double box with six off-shell legs in momentum space):
\begin{align} \label{crossdbl}
\ket{\text{double-cross}} = \int \frac{\mathrm{d}^4 x_0 \mathrm{d}^4 x_{0'}}{x_{10}^2 x_{50}^2 x_{60}^2 x_{00'}^2 x_{20'}^2 x_{30'}^2 x_{40'}^2}\,.
\end{align}
It is an eigenfunction of the six-point monodromy
\begin{align} \label{cross2ef}
L_6[4,5]L_5[3,4]L_4[3,4]L_3[2,3]L_2[1,2]L_1[1,2] \,\ket{\text{double-cross} } = [3]^2[4]^3[5]\,\ket{\text{double-cross}}\idop\,.
\end{align}
An explicit expression for this integral in terms of familiar special functions is not yet known. It is believed to be given by a class of elliptic functions.

The proof is given by the series of transformations shown in Fig.~\ref{figdblcrss}.
Similarly to the cross integral, we use eq.\ (\ref{vacuum}) to extend the monodromy (the first transformation in Fig.~\ref{figdblcrss})
by two additional Lax operators corresponding to the integration points $x_0$ and $x_{0'}$: 
\begin{align} \label{8ptmonodromy}
L_6[4,5]L_5[3,4]L_4[3,4]L_3[2,3]L_2[1,2]L_{0'}[2,0]L_1[1,2]L_0[2,0]\,.
\end{align} 
Then we show that the integrand of eq.\ (\ref{crossdbl}) is an eigenfunction of the eight-point monodromy (\ref{8ptmonodromy}).
We pull the monodromy through the scalar propagators and repeatedly use eqs.\ (\ref{vacuum}) and (\ref{intw}).
Integrating the two auxiliary sites $L_0^T[2,0] \cdot 1 \sim \idop$ and $L_{0'}^T[2,0] \cdot 1 \sim \idop$ by parts, cf.\ eq.\ (\ref{vacuum}), we come back to the six-point monodromy and we obtain the eigenvalue relation (\ref{cross2ef}).

\subsection{Multi-Loop Integrals} \label{MLI}
 
\begin{figure}
\begin{center}
\includegraphics[scale =1]{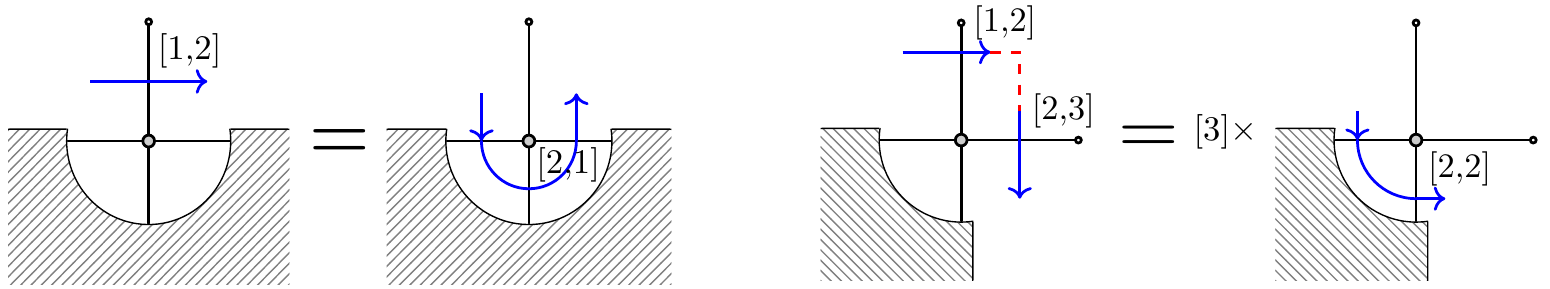}
\end{center}
\caption{\label{elmtrns} Local transformations of the monodromy contour.  The dashed part represents the rest of the graph which is not touched by the current transformation.  
Pushing the contour inside the graph involves integrations by parts. If the initial monodromy acts on an $L$-loop graph, then, after one local transformation, the new monodromy (transformed contour) acts on the $(L-1)$-loop integral.}
\end{figure}

\begin{figure}
\begin{center}
\includegraphics[scale=1]{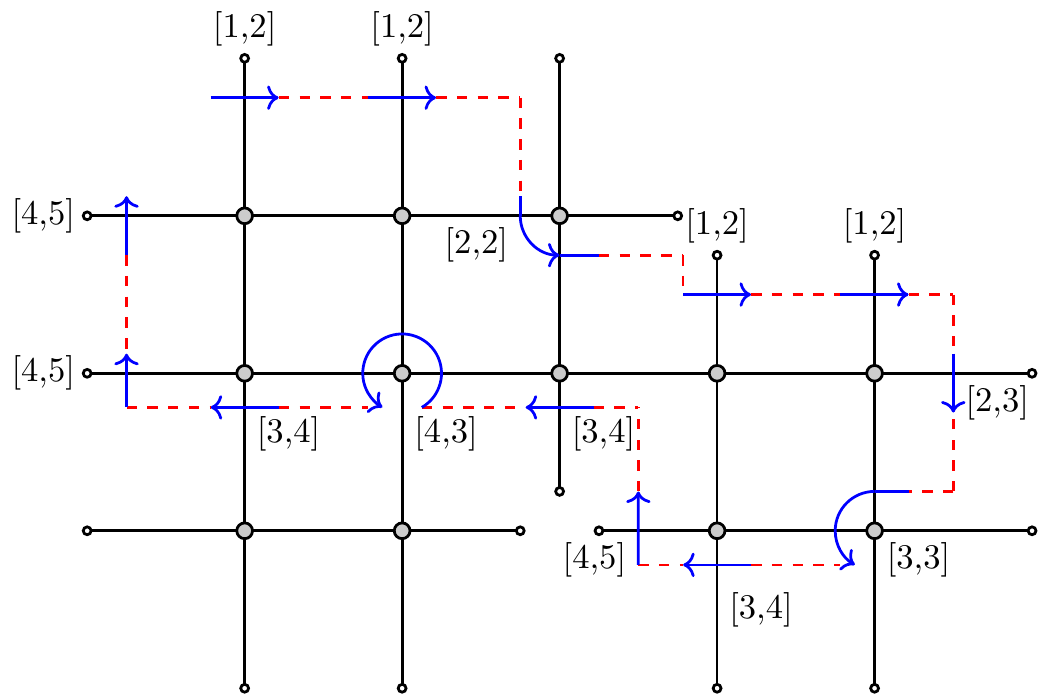}
\end{center}
\caption{\label{Fig4} We deform the contour $\cal C$ inside the graph. Algebraically this means that we change the monodromy. Blue segments denote Lax operators in the product. We see that we can always apply the transformations of the  Fig.~\ref{elmtrns}  to any part of this intermediate contour and thus shrink it further eventually reducing it to a point (i.e.\ the identity matrix). }
\end{figure}

Now we want to consider a generic fishnet graph, reproducing all possible features of boundary geometries. As a representative of such generic graphs, we can take the one  depicted    in Fig.~\ref{Fig2}.
 The monodromy around the boundary of this graph, the ``lasso'', is presented in Fig.~\ref{Fig3}. We interpret the fishnet graph as an intertwiner for the monodromy matrix, since it is built from scalar propagators which are intertwiners
according to eq.\ (\ref{intw}). So, according to our ``lasso'' method, we want to pull the $2M$-point fishnet graph through the $2M$-site monodromy.
We prefer to explain the proof of eq.\ (\ref{main}) using pictures. Graphically, we pull the oriented contour $\cal C$ through the graph.

To deform the monodromy contour we use the elementary transformations in Fig.~\ref{elmtrns}. One can justify them using the same manipulations as in Fig.~\ref{crossfig} for the single cross integral. Now, using these transformations, we consecutively pull the monodromy contour in Fig.~\ref{Fig3} through the graph using the transformations in Fig.~\ref{elmtrns}. An intermediate step is shown in Fig.~\ref{Fig4}. We also use the intertwining relations (\ref{intw})
and the local pseudo-vacuum of eq.\ (\ref{vacuum}), to replace the Lax operator $L_k[i,i+2]$ by the diagonal matrix $[i+2] \, \idop$. In this way we shrink the monodromy contour and effectively decrease the length of the non-compact spin chain represented by the monodromy. Finally the contour shrinks to a point, so that the monodromy is proportional to the identity matrix. Hence, relation (\ref{main}) is proven.

\section{Yangian Symmetry of  Bi-Scalar Fishnet Graphs: Irregular Boundary} \label{irreg}

\begin{figure}
\begin{center}
\includegraphics[scale = 1.7]{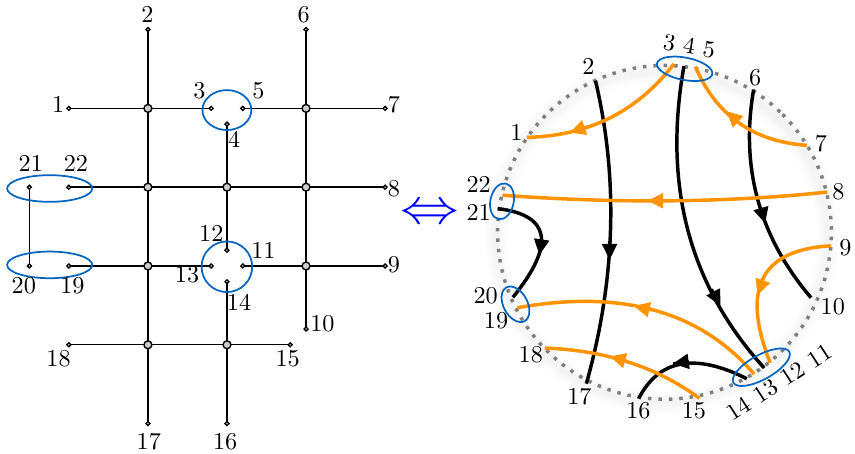}
\end{center}
\caption{ \label{Fig-join} To obtain a correlator diagram with irregular boundary we identify external points of a regular boundary  ($2M = 22$). This corresponds to identifications of adjacent coordinates $x_3 = x_4 =x_5$, $x_{11} = x_{12} = x_{13} = x_{14}$, $x_{19} = x_{20}$ and $x_{21}=x_{22}$  in the correlator $K$, cf.\ eq. (\ref{TrCorrelator}). In the picture on the right-hand side, 
the flavors of fields are explicitly indicated. 
We are allowed to identify coordinates of only those adjacent fields in $K$ which are ordered as in the interaction vertex $\phi^\dagger_1 \phi^\dagger_2 \phi_1 \phi_2$, see eq. (\ref{Lthree}), up to cyclic shifts. No further junctions are allowed in this picture.}
\end{figure}

\begin{figure}
\begin{center}
\includegraphics[scale=1]{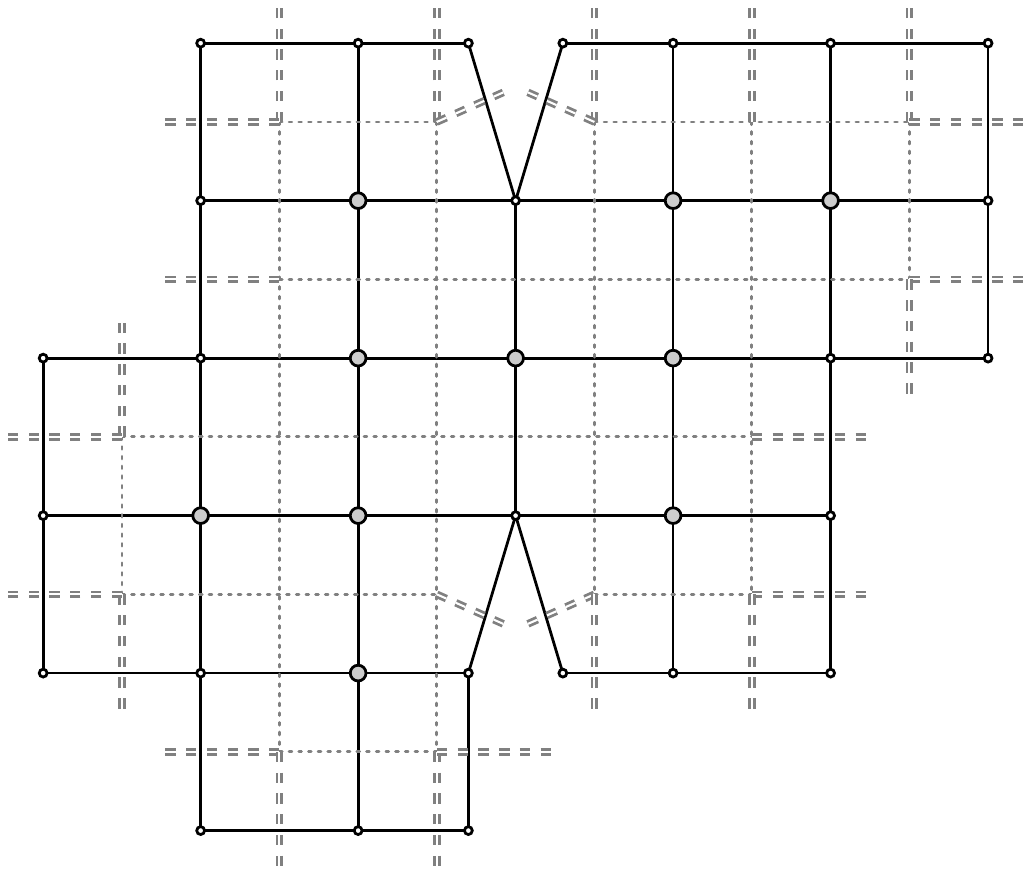}
\end{center}
\caption{ \label{Fig7} A fishnet correlator graph with a generic boundary is drawn by solid lines. Here several propagators meet in external points. Integration points are denoted by filled blobs. Propagators stretched between a pair of external vertices are inert with respect to loop integrations. The dual (momentum-space) graph is drawn by dotted lines. All incoming momenta are off shell and denoted by double lines. 
}
\end{figure}

Now we want to include into our considerations the coordinate space fishnet graphs with more generic boundaries. These are the fishnet graphs where some of the neighboring external legs are joined into one point. They correspond to correlator diagrams for $K(x_1,\ldots,x_{2M})$ in eq.\ (\ref{TrCorrelator}), where several neighboring external points $x_1,\ldots,x_{2M}$ coincide. 

To be more specific, let us consider a regular diagram decorated with flavors, see Fig.~\ref{Fig:biscalar} and Fig.~\ref{Fig-join}. There is a one-to-one correspondence between a diagram and the ordering (up to cyclic shifts) of scalar fields $\chi_j(x_j), \, j = 1 , \ldots,2M$ which carry one of the four flavors. We want to identify  coordinates of some of the adjacent fields. Putting several fields in a common space-time point might cause divergences in loop diagrams. However in some cases this junction results in finite loop integrals. Let us consider the product of bi-scalar fields of length $4L$
\begin{align*}
\ldots\phi^\dagger_1 \phi^\dagger_2 \phi_1 \phi_2 \phi^\dagger_1 
\underbrace{\phi^\dagger_2 \phi_1 \phi_2 \ldots \phi^\dagger_1 \phi^\dagger_2 }\phi_1 \phi_2\ldots\,,
\end{align*}
where the fields are ordered as in the interaction vertex of the Lagrangian (\ref{Lthree}). We cut a piece of arbitrary length out of this product (indicated by the curly bracket in the previous formula). 
We allow only this type of products of fields in $K$ to be put at a common space-time point because they correspond to a vertex vertex cut out of the original lattice. 
An example is given in Fig.~\ref{Fig-join}, where we joined two, three and four neighboring points: 
$\phi_1 \phi_2$ sits in $x_{19} = x_{20}$;
$\phi_2^\dagger \phi_1$ sits in $x_{21} = x_{22}$; 
$\phi_1^\dagger \phi_2^\dagger \phi_1$ sits in $x_{3} = x_{4} = x_{5}$; 
$\phi_1 \phi_2\phi_1^\dagger \phi_2^\dagger$ sits in $x_{11} = x_{12} = x_{13} = x_{14} $.
The external points $x^\mu_i$ in the left  figure are decorated by small white blobs as opposed to the filled interior vertices (integration points). We can consider the junction of an arbitrary number of fields. The graphs with junction of five or more fields live on a multi-sheet lattice, such as in Fig.~\ref{FigOverlap}.
The number of external points in a graph with irregular boundary does not have to be even, so we denote the set of external points by $x_1,\ldots,x_{N}$.

Let us stress that we do not allow for arbitrary junctions of neighboring legs, since the corresponding integrals 
are  potentially IR divergent.
Another, more complicated example of the graphs we consider is given in Fig.~\ref{Fig7}, along with its dual momentum-space graph. 
As we can see, it corresponds to a loop integral with all inflowing momenta off shell.
Note that dashed double lines in Fig.~\ref{Fig7} correspond to trivial momentum-space propagator factors, involving only external fixed momenta. However we will need them in the following.
They correspond to coordinate space propagators stretched between two external vertices which are not involved into the integration. 
Note that the graph in Fig.~\ref{Fig7} originates from  a  singular square lattice (see Footnote~\ref{ftnt3} and  Fig.~\ref{FigOverlap}): the plaquettes around the vertices with five neighbors there cannot be cut out of a usual regular square lattice. Such a lattice should have a conical singularity with the angle of a cone equal to \(\frac{5}{2}\pi\). We recall that such conical singularities can appear only at the boundary of the disc. 

\begin{figure}
\begin{center}
\includegraphics[scale = 1]{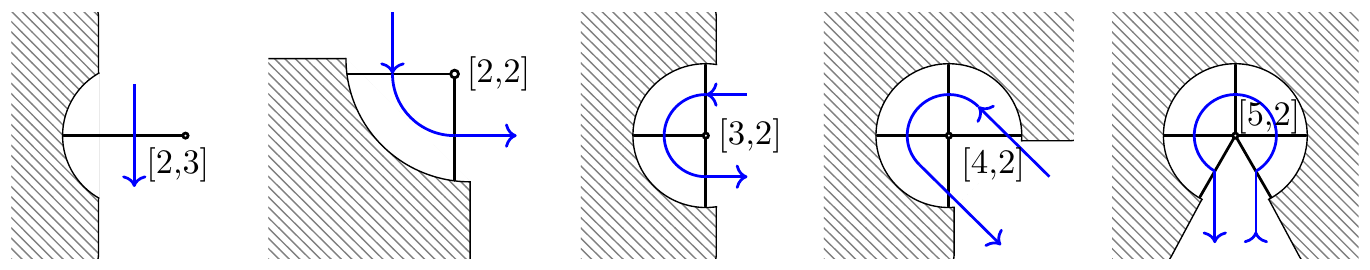}
\end{center}
\caption{ \label{Fig8} 
The Lax operator from the monodromy acting onto the junction of: (A) one, (B) two, (C) three, (D) four, (E) five external legs. Rotating this picture by $\pm \pi/2$ we shift both inhomogeneities $\delta \to \delta \pm 1$.}
\end{figure}

\begin{figure}
\begin{center}
\includegraphics[scale=1]{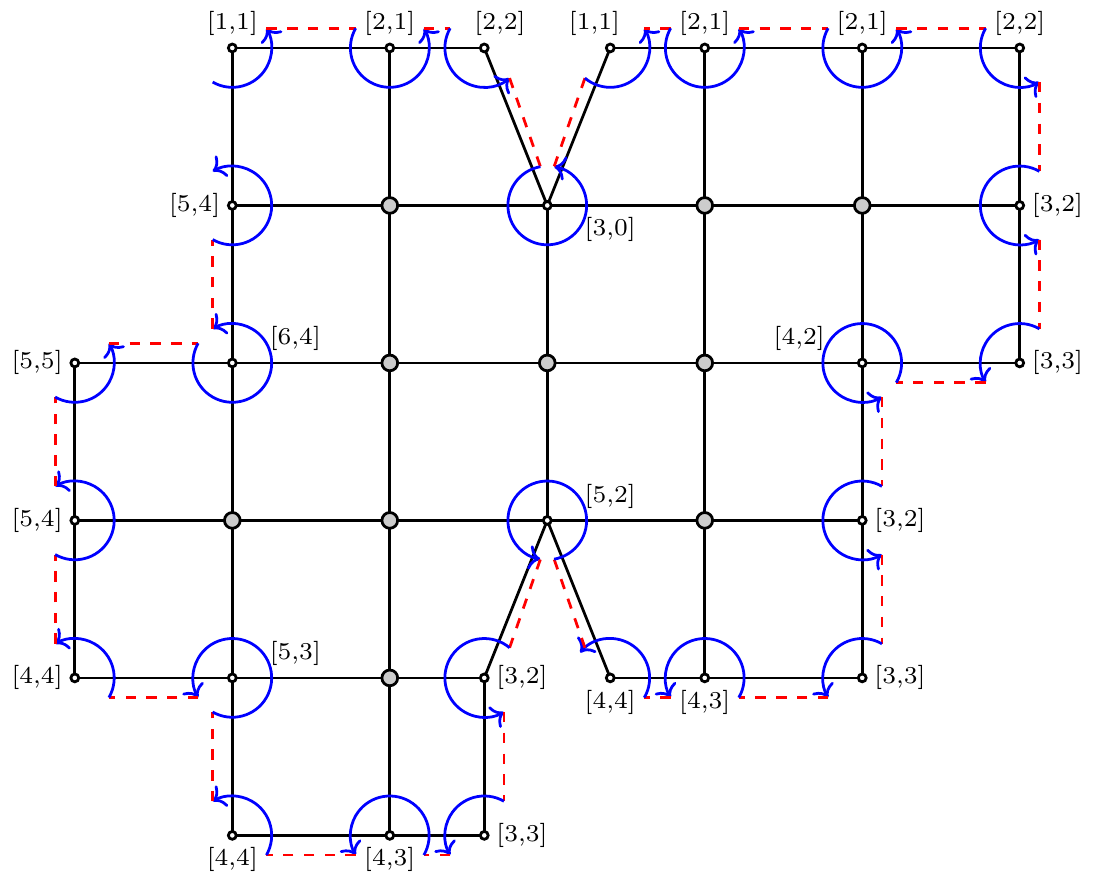}
\end{center}
\caption{ \label{Fig10} The fishnet graph with irregular boundary depicted in Fig.~\ref{Fig7} is a Yangian invariant, i.e.\
it is an eigenfunction of the monodromy with the indicated inhomogeneities $[\delta^+_i,\delta^-_i]$. 
The assignment of inhomogeneities follows the rule in Fig.~\ref{Fig8}.
}
\end{figure}

Let us now act with the monodromy on a correlator diagram with irregular boundary. 
The monodoromy contour~$\cal C$ has to cross all external points.
We need to specify how we depict a Lax operator which acts simultaneously onto a junction of several external legs. 
We draw the monodromy contour crossing one, two, three, etc.\ legs as in Fig.~\ref{Fig8}. 
The rule for the assignment of inhomogeneities to Lax operators is given there. The application of this rule to the correlator diagram from Fig.~\ref{Fig7} is given in Fig.~\ref{Fig10}. The external vertices with junction of five or more propagators appear
in graphs living on a  singular square  lattice (with conical singularities). The junction of external legs changes the conformal weight $\Delta$ (we have $\Delta = k$ for the junction of $k$ legs, and consequently $\delta^+ -\delta^- + 2 =k$). Hence, the inhomogeneities of Lax operators, see eq.\ (\ref{u+u-}), differ from those for a regular boundary. So considering irregular boundaries, we deal with a spin chain carrying different principal series representations of $\alg{so}(2,4)$ on different sites.

A fishnet graph $\ket{G}$ with a boundary consisting of the elements from Fig.~\ref{Fig8} (and their rotations by a multiple of $\pm \pi/2$) is a Yangian invariant
\begin{align} \label{main2}
\left(\prod_{i \in {\cal C}} L_i[\delta_{i}^{+},\delta_{i}^{-}]  \right) \ket{G} = 
\lambda(\vec u)\,\ket{G}  \idop \,.
\end{align}
The rule for assigning the inhomogeneities $\delta^\pm_i$ is given in Fig.~\ref{Fig8}. 
An example graph is given in Fig.~\ref{Fig10}. One can establish this relation following the line of the proof for the regular boundary, Section~\ref{MLI}. So we pull the monodromy through the diagram using a sequence of elementary transformations given in Fig.~\ref{elmtrns}. 
Note that we have already encountered irregular boundaries in Section~\ref{MLI} at the intermediate steps of the shrinking contour, see Fig.~\ref{Fig4}, where one can see three neighbors at the same point at its intermediate boundary (like the vertex carrying the Lax operator with argument \([4,3])\).

Explicit expressions for $\lambda$ can be worked out for each particular graph. We just need to keep track of numerical factors pulling the monodromy contour through the graph. There is, however, a quicker way to find $\lambda$. Applying $N$ times the cyclicity property of eq.\ (\ref{cyclT}), we obtain the initial eigenvalue problem with spectral parameter $u$ shifted uniformly in all Lax operators. Comparing the eigenvalue of the initial eigenvalue problem and of the cyclically rotated one, we conclude that they satisfy the finite-difference equation
\begin{align} \label{BAE}
\frac{\lambda(u)}{\lambda(u-4)} = \frac{P(u)}{P(u-2)} ,
\end{align}
where $\lambda(u)$ is a polynomial of degree $N$, and $P(u)$ is a known (for each contour) polynomial of degree $2N$,
\begin{align}
\lambda(u) = \prod_{i= 1}^N (u + a_i) \;,\;\;\;\quad  P(u) = \prod_{i=1}^N (u + \delta^+_i)(u + \delta^-_i).
\end{align}
Eq.\ (\ref{BAE}) is reminiscent of the Bethe Ansatz Equations (BAEs). In comparison with the usual BAEs, it is easy to solve equation (\ref{BAE}), i.e.\ to find roots of the polynomial $\lambda(u)$.
This equation implies that the inhomogeneities cannot be arbitrary. Indeed, the sets $\{\delta^{\pm}_i\}_{i=1}^N$ and $\{\delta^{\pm}_i - 2\}_{i=1}^N$ have to contain at least $N$ common elements.
Then the parameters $\{a_i\}_{i=1}^N$ determining the eigenvalue $\lambda(u)$ are such that $\{a_i,\delta^{\pm}_i - 2\}_{i=1}^N = \{a_i - 4,\delta^{\pm}_i \}_{i=1}^N$. Solving eq. (\ref{BAE}) in the example in Fig.~\ref{Fig10}, we immediately find $\lambda(u) = [2] [3]^7 [4]^{11} [5]^6 [6]$.

Alternatively, we may perturbatively solve the relation \eqref{BAE} for the coefficients in the $u$-expansion of the monodromy eigenvalue $\lambda$. This will be helpful for making connection to the first realization of the Yangian in Section \ref{sec:firstreal}. For the first two orders of the expansion we obtain
\begin{align}\label{eq:expansionRHSfirst}
\lambda(\vec{u})=
u^n 
&+\half u^{n-1}  \sum_{k=1}^n \hat \delta_k
+\quarter u^{n-2} \Big[\sum_{i<j=1}^n \hat \delta_i \hat \delta_j -\sfrac{1}{2}\sum_{j=1}^n\hat \Delta_j  \Big]
+\mathcal{O}(u^{n-3})
\end{align}
In Appendix \ref{sec:eigenexp} we display the first five orders of the expansion.
Here we use the shorthand notations $\hat \delta_k=\delta_k^++\delta_k^-+2$ and $\hat\Delta_i=\Delta_i(\Delta_i-4) $ with  $\Delta_k=\delta_k^+-\delta_k^-+2$.

\section{Scattering Amplitudes and Cuts of Fishnet Graphs}

Now we want to consider scattering amplitudes in the bi-scalar theory with massless external states.
More generally, we want to put some of the inflowing momenta in the momentum Feynman integrals considered above on shell, like in Fig.~\ref{Fig7}. We use region momenta $x_i^{\mu}$ as the amplitude variables. They are coordinates of the external legs of fishnet graphs. All variables $x^\mu_i$ were independent and unconstrained in the previous considerations. However for a light-like momentum, $p_i^2 = 0$, the region momenta have to be constrained by $x_{i \,i+1}^2 = 0$. 

Fortunately, we can easily impose the constraints on region momenta in a way consistent with Yangian symmetry.
We note that scalar propagators $x_{ij}^{-2}$ and distributions $\delta(x_{ij}^2)$ have the same conformal weights
and they satisfy identical intertwining relations --- compare eqs. (\ref{intw}) and (\ref{intwdelt}). So if we take a fishnet graph $\ket{G}$ which respects the Yangian symmetry, eq.\ (\ref{main2}), and replace a scalar propagator $x_{ij}^{-2}$  by the delta-function $\delta(x_{ij}^2)$, then we obtain another Yangian invariant $\ket{G}_{\text{cut}}$ which satisfies the same eigenvalue relation (\ref{main2}) with the same inhomogeneities:
\begin{align} \label{cut}
T(\vec u) \, \ket{G} = \lambda(\vec{u})\, \ket{G} \idop  \;\;\;\Rightarrow \;\;\;  T(\vec u) \, \ket{G}_{\text{cut}} = \lambda(\vec{u})\, \ket{G}_{\text{cut}} \idop\,.
\end{align}
In the literature, this substitution of a number of propagators $1/(p^2+i \epsilon)$ in a Feynman graph by their imaginary parts $\delta(p^2)$ is known as a generalized cut. So the cutting of Feynman graphs is consistent with Yangian symmetry.

\begin{figure}
\begin{center}
\includegraphics[scale=1]{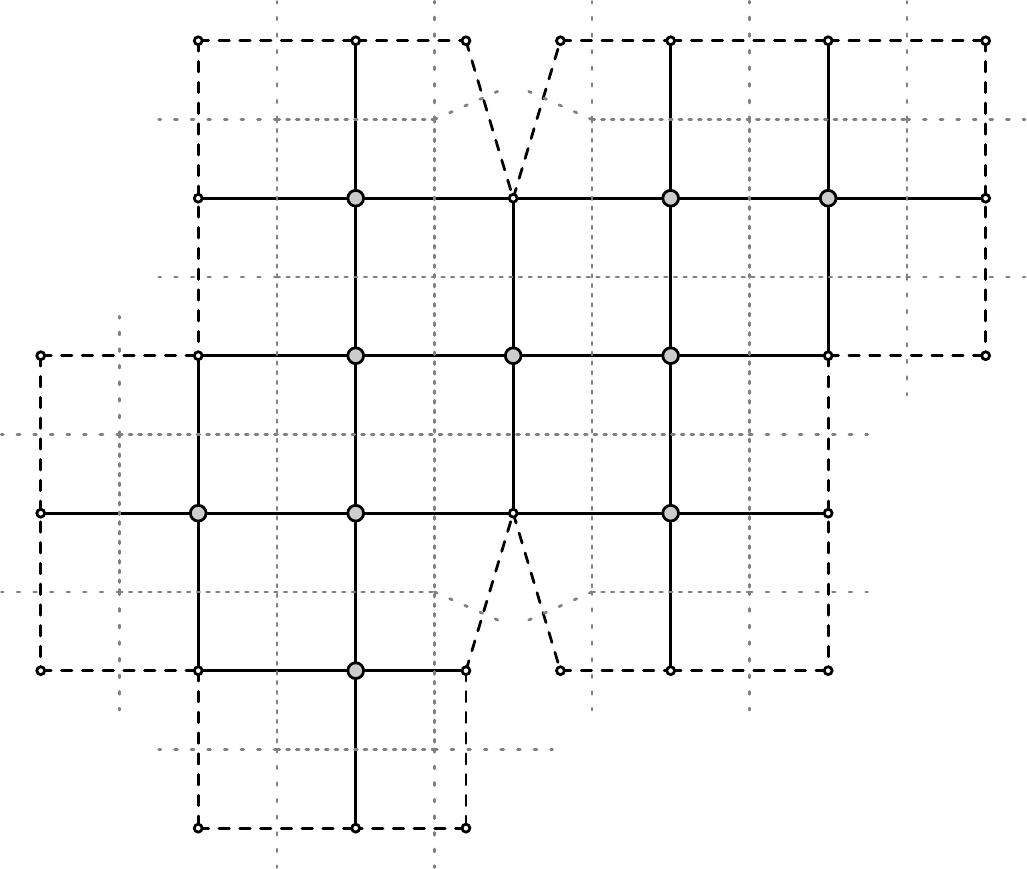}
\end{center}
\caption{ \label{Fig11} A fishnet correlator graph and its dual which represents a genuine scattering amplitude in the bi-scalar theory.
The fishnet graph is formed by solid lines (which denote scalar propagators $x_{ij}^{-2}$) and dashed lines (which denote cut propagators $\delta(x_{ij}^2)$). The dual graph is drawn by dotted lines which denote scalar propagators $p_i^{-2}$. All inflowing momenta are on shell. They are denoted by thin dotted lines which cross cut propagators. This picture is obtained from Fig.~\ref{Fig7} by cutting a number of (or all) propagators along the boundary in the correlator graph. This cutting in $x$-space corresponds to the LSZ amputation of external propagators in the dual $p$-space graph.
}
\end{figure}

Now, if we cut a propagator stretched between two adjacent external points $x_i$ and $x_{i+1}$ of a fishnet graph, we put the corresponding momentum $p_i$ of the dual graph on shell, i.e.\ we set $p_i^2=0$. For example, in the off-shell Feynman graph in Fig.~\ref{Fig7} this cutting procedure
leads to the Feynman integral in Fig.~\ref{Fig11} with all inflowing momenta light-like. 
So we describe the amplitudes using region momenta $x_i^{\mu}$ and consider them as distributions, 
$\ket{G}_{\text{cut}^n} = \prod_{i=1}^n \delta(x_{i \,i+1}^2) \cdot {\cal A}_n$, where ${\cal A}_n$ is a regular function.
We can also consider mixed objects, where only a part of the inflowing momenta are on shell. We just need to cut a smaller number of propagators in Fig.~\ref{Fig7}, which also results in distributions.

One can see that for any scattering amplitude in the bi-scalar theory the corresponding dual fishnet graph has the topology considered in Sections \ref{sec:YangianSymmetry} and \ref{irreg}.

The loop amplitude diagrams for massless particles do not suffer from IR or collinear divergences. Such divergences are typical for loop diagrams with emission of a massless particle through a cubic vertex. However, in our diagrams the external light-like momenta flow into quartic vertices, and in the corners light-like momenta enter in pairs (each of which is equivalent to one off-shell momentum).\footnote{We are grateful to G.~Korchemsky and J.~Henn for discussions on this point.
}

According to eq.\ (\ref{cut}) we are allowed to cut internal propagators as well. In this way we localize some of the loop integrations reducing the complexity of the Feynman integral. Let us give several examples. 
Cutting four propagators of the cross integral \eqref{cross} we localize all integrations and the result is an algebraic function $1/\sqrt{\det g}$, where $g_{ij} = x_{ij}^2$, $i,j=1,\ldots,4$, \cite{Drummond:2013nda}. 
It is a Yangian invariant, i.e.\ it satisfies the same eigenvalue relation (\ref{crossef}) as the cross integral. We checked this independently. Let us note that the solution $x_0$ of the constraints $x_{10}^2 = x_{20}^2 = x_{30}^2 = x_{40}^2 = 0$ is complex-valued. So strictly speaking it does not belong to the support of the delta-function. In fact, we need to understand the cutting as deforming the integration contour and taking the residue in the pole specified by the cut propagator.
Cutting all seven propagators of the double box with six external legs, we localize seven of eight Feynman integrations. The resulting cut integral has a one-fold integral representation, which can be expressed in terms of elliptic integrals. The cut of the double-cross integral, as well as the integral itself, satisfies the eigenvalue relation (\ref{cross2ef}), and so it is a Yangian invariant.

\section{Fishnet Graphs and First Realization of the Yangian}
\label{sec:firstreal}

Above we have formulated the Yangian symmetry of fishnet graphs in the language of the RTT realization. In order to study explicit constraints in form of (non-local) differential equations in the future, it should be useful to make contact to Drinfel'd's first realization of the Yangian algebra \cite{Drinfeld:1985rx}.
 This first realization is closer to the formulation of Lie algebras and given in terms of two types of generators, which can be obtained from the expansion of the monodromy $T(u)$.

\subsection{First Realization}

In general, the Yangian is an infinite-dimensional algebra which can be understood as an extension of an underlying Lie algebra. 
In its first realization, it is spanned by the Lie algebra generators $\gen{J}^A$ and a second set of generators $\widehat{\gen{J}}^A$ transforming in the adjoint representation of the underlying Lie algebra, i.e.\
\begin{align}
\bigl[ \gen{J}^A ,\gen{J}^B\bigr]=f^{AB}{}_{C} \, \gen{J}^C \hspace{2cm} \bigl[ \gen{J}^A ,\widehat{\gen{J}}^B \bigr]=f^{AB}{}_{C} \, \widehat{\gen{J}}^C \, .
\label{eqn:Yangiancomrel}
\end{align}
The generators $\widehat{\gen{J}}^A$ are typically referred to as the level-one generators while the Lie algebra generators, in the context of the Yangian, are called the level-zero generators. In addition to the two Jacobi-identities 
\begin{align}
\bigl[\gen{J}^{A}, \bigl[ \gen{J}^B ,\gen{J}^{C}\bigr] \bigr]+\mbox{cyclic}= 0, \hspace{2cm} \bigl[\gen{J}^{A}, \bigl[ \gen{J}^B ,\widehat{\gen{J}}^{C}\bigr] \bigr]+\mbox{cyclic}= 0 ,
\end{align}
the level-zero and level-one generators should obey the Serre relations
\begin{align}
\bigl[\widehat{\gen{J}}^{A}, \bigl[ \widehat{\gen{J}}^B ,\gen{J}^{C}\bigr] \bigr]+\mbox{cyclic}= \tfrac{1}{4} f^{AG}{}_{D} f^{BH}{}_{E} f^{CK}{}_{F} f_{GHK} \gen{J}^{\{D} \gen{J}^{E} \gen{J}^{F\}} \, ,
\end{align}
where the symbols $f_{GHK}$ follow from $f^{GH}{}_K$ by lowering two indices with the inverse of the Cartan--Killing form of the underlying Lie algebra. The brackets $\{ \quad \}$  denote total symmetrization of the enclosed indices. From an algebraic point of view the Yangian is a Hopf algebra and as such it comes equipped with a coproduct. The coproduct furnishes a prescription for how to extend a single-site representation to a multi-site representation and takes the following standard form
\begin{align}
&\Delta(\gen{J}^A)=\gen{J}^A \otimes \idop + \idop \otimes \gen{J}^A  \nonumber \\
&\Delta( \widehat{\gen{J}}^A)= \widehat{\gen{J}}^A \otimes \idop + \idop \otimes  \widehat{\gen{J}}^A + \tfrac{1}{2}  f^{A}{}_{BC} \,\gen{J}^C \otimes \gen{J}^B \, .
\end{align}
Based on this coproduct we can write down the following ansatz for a general level-one generator
\begin{align}
\gen{\widehat J}^{A}=\tfrac{1}{2} f^A{}_{BC} \sum_{j<k} \gen{J}^C_j \gen{J}^B_k +\sum_{k} v_k \gen{J}^{A}_k \, .
\label{eqn:generallevone}
\end{align}
Here, $f^A{}_{BC}$ are the inverse structure constants which follow from the ordinary structure constants by lowering one Lie algebra index with the Cartan--Killing form. The so-called evaluation parameters $v_k$ can be chosen arbitrarily without spoiling the above defining relations of the Yangian. In many cases, like for example in the context of scattering amplitudes in $\mathcal{N}=4$ SYM theory \cite{Drummond:2009fd}, the local contribution (i.e. the single sum term) is absent and we have $v_k=0$ for all $k$.

\subsection{Pedagogic Examples: Cross and Double Cross}

In order to illustrate the explicit form of the above Yangian generators on fishnet graphs, we consider two simple examples. That is, we explicitly construct the level-one generators, which annihilate the one-loop cross and the two-loop double cross Feynman graph.

\paragraph{Cross.}

The cross diagram corresponds to the following Feynman integral:
\begin{align}
\ket{\text{cross}} =   \int   \frac{\mathrm{d}^4 x_{0}}{x_{10}^2x_{20}^2x_{30}^2x_{40}^2} \, .
\label{eqn:boxintegral}
\end{align}
Here, we again consider the four external coordinates $x_1,\ldots, x_4$ as independent, i.e.\ we interpret the cross graph as an off-shell correlator diagram.  
As discussed in the previous sections, the integral \eqref{eqn:boxintegral} is invariant under the generators of the four-dimensional conformal algebra $\mathfrak{so}(2,4)$ with generators as defined in equation \eqref{eqn:diffrepconfgen}: 
\begin{align}
\gen{J}^A \, \ket{\text{cross}}  =0 \, .
\label{eqn:Boxconfsym}
\end{align}  
Here $\gen{J}^A$ is defined as
\begin{align}
\gen{J}^A = \sum\limits_{i=1}^{4} \gen{J}_i^A \, ,
\end{align}
with $\gen{J}_i^A \in \{\gen{P}_{i,\mu},\gen{L}_{i,\mu \nu},\gen{D}_i, \gen{K}_{i,\mu}\}$ being the single-site generators of the conformal algebra \eqref{eqn:diffrepconfgen} with the conformal dimensions $\Delta_i$ fixed uniformly to be equal to $1$.  
We will now demonstrate how to construct level-one generators which annihilate the cross integral. 
Note that in the case of the conformal algebra it suffices to show invariance under one level-one generator. The level-zero invariance together with the Yangian commutation relations \eqref{eqn:Yangiancomrel} guarantees that all the other level-one generators annihilate the cross integral as well. 
In what follows, we will choose the simplest level-one generator which is the level-one momentum generator $\gen{\widehat P}^{\mu}$. Using the formula \eqref{eqn:generallevone} we find the following expression for the bi-local piece of $\gen{\widehat P}^{\mu}$:
\begin{align}
\gen{\widehat P}^{\mu}_{\mathrm{bi}}=-\sfrac{i}{2} \sum_{j<k=1}^n \bigl[ (\gen{L}_j^{\mu \nu} + \eta^{\mu \nu} \gen{D}_j)\gen{P}_{k,\nu}-(j \leftrightarrow k)\bigr]\, ,
\label{eqn:Plevel1bi}
\end{align}
where $\gen{L}_j^{\mu \nu}$, $\gen{D}_j$ and $\gen{P}^{\mu}_{k}$ are the single-site conformal generators as introduced above. Applying this generator to the box integral \eqref{eqn:boxintegral} yields
\begin{align}
\gen{\widehat P}^{\mu}_{\mathrm{bi}} \, \ket{\text{cross}}  = \bigl(\gen{P}^{\mu}_2+2\gen{P}^{\mu}_3+3\gen{P}^{\mu}_4 \bigr) \, \ket{\text{cross}}\, .
\label{eqn:Plevel1oncross}
\end{align}
The form of the right hand side of the above equation makes it obvious that we can use the freedom to choose the $v_k$'s in equation \eqref{eqn:generallevone} to construct a true symmetry generator. Explicitly, we define the full level-one momentum generator as
\begin{align}
\gen{\widehat P}^{\mu}_{\mathrm{cr}}:= \gen{\widehat P}^{\mu}_{\mathrm{bi}} - \gen{P}^{\mu}_2- 2\gen{P}^{\mu}_3-3\gen{P}^{\mu}_4 \, .
\end{align}

\paragraph{Double Cross.}

As a second example, let us consider the double cross diagram. The corresponding Feynman integral reads
\begin{align}
\ket{\text{double-cross}}=\int  \frac{\mathrm{d}^4 x_{0} \, \mathrm{d}^4 x_{0'}}{x_{10}^2x_{20}^2x_{30}^2 x_{40'}^2 x_{50'}^2 x_{60'}^2 x_{00'}^2} \, .
\label{eqn:doublecrossint}
\end{align}
Applying the bi-local generator \eqref{eqn:Plevel1bi} to the double cross integral \eqref{eqn:doublecrossint} yields
\begin{align}
\gen{\widehat P}^{\mu}_{\mathrm{bi}} \, \ket{\text{double-cross}} =  \bigl( \gen{P}^{\mu}_2 + 2\gen{P}^{\mu}_3+2\gen{P}^{\mu}_4+3\gen{P}^{\mu}_5+4\gen{P}^{\mu}_6 \bigr) \, \ket{\text{double-cross}} \, .
\end{align}
Again, we see that we can define an algebraically consistent level-one momentum generator that annihilates the integral \eqref{eqn:doublecrossint} by choosing the inhomogeneities as follows: 
\begin{align}
\gen{\widehat P}^{\mu}_{\mathrm{dcr}}:= \gen{\widehat P}^{\mu}_{\mathrm{bi}} - \gen{P}^{\mu}_2- 2\gen{P}^{\mu}_3-2\gen{P}^{\mu}_4-3\gen{P}^{\mu}_5-4\gen{P}^{\mu}_6 \, .
\end{align}
Given the discussion of fishnet graphs within the RTT realization in the previous sections, it is in fact clear that the above procedure has to work for an arbitrary graph of fishnet type. 
This is due to the fact that the monodromy matrix packages all Yangian generators in a very efficient way, cf.\ \eqref{TJ}. For this reason, we will now derive in detail the generic relation between the monodromy matrix, the evaluation parameters $v_k$ and the explicit level-one generators.

\subsection{Monodromy Expansion}

In order to obtain the explicit form of the Yangian generators, we expand the following monodromy matrix in the spectral parameter $u$:
\begin{equation}
T(\vec{u})
=L_n(u+\delta_n^+,u+\delta_n^-)L_{n-1}(u+\delta_{n-1}^+,u+\delta_{n-1}^-)\dots L_{1}(u+\delta_1^+,u+\delta_1^-)
\end{equation}
We employ the Lax operator given in \eqref{eq:Lax1} and \eqref{Lax}, which yields
\begin{align}\label{eq:monexp1}
T(\vec{u})
=&
u^n \idop
+\half u^{n-1}\sum_{k=1}^n\big( \hat \delta_k\idop+  \rho_k^{ab} s_{ab}\big)
\\
&+\sfrac{1}{8} 
u^{n-2}\bigg[\sum_{k=1}^n\sum_{j=1}^{k-1}+\sum_{j=1}^n\sum_{k=1}^{j-1}\bigg]\big( \hat \delta_j\idop+  \rho_j^{ab} s_{ab}\big)\big( \hat \delta_k\idop+  \rho_k^{cd} s_{cd}\big)
+\dots.
\nonumber
\end{align}
Here we use the shorthand notation $\hat \delta_k=\delta_k^++\delta_k^-+2$ and the abbreviations of Section~\ref{sec:conf_Lax}:  $s_{ab} = s(M_{ab})$ and $\rho_{k,ab}=\rho_{\Delta_k}(M_{ab})$. In order to identify the level-zero and level-one generators at order $u^{n-1}$ and $u^{n-2}$ of the above expansion, respectively, we also have to consider the function on the right hand side of the generic monodromy equation \eqref{main2}, whose expansion is given in \eqref{eq:expansionRHSfirst} (cf.\ \eqref{main} for the case $\Delta=1$):
\begin{align}\label{eq:expansionRHS}
\lambda(\vec{u})=
&u^n 
+\half u^{n-1}  \sum_{k=1}^n \hat \delta_k 
+\quarter u^{n-2} \Big[\sum_{i<j=1}^n \hat \delta_i \hat \delta_j -\sfrac{1}{2}\sum_{j=1}^n\hat \Delta_j  \Big]
+\mathcal{O}(u^{n-3}).
\end{align}
Here $\hat\Delta_i=\Delta_i(\Delta_i-4) $ with $\Delta_k=\delta_k^+-\delta_k^-+2$.
We can thus subtract the eigenvalue $\lambda$ from the monodromy to find the following operator which annihilates invariants under the Yangian algebra for arbitrary spectral parameter $u$:
\begin{align}\label{eq:moneqexp}
T(\vec{u})&-\lambda(\vec{u})\idop=
 0\times u^n \idop
+ u^{n-1} \Big[\half \sum_{k=1}^n \rho_k^{ab} s_{ab} \Big] 
\\
&+u^{n-2}\Big[\quarter\sum_{j<k=1}^n \rho_k^{ab} \rho_j^{cd}s_{ab}s_{cd}+\quarter\sum_{k=1}^n\mathop{\sum_{j=1}}_{j\neq k}^n\hat \delta_j \rho_k^{ab}s_{ab}-\sfrac{1}{8}\sum_{k=1}^n(4-\Delta_k)\Delta_k  \idop  \Big]+\dots.
\nonumber
\end{align}
This allows to identify the level-zero generator 
\begin{equation}\label{eq:levzfrommon}
\gen{J}^{ab}=\half \sum_{k=1}^n \rho_k^{ab}
\end{equation}
at order $u^{n-1}$ as a symmetry operator which annihilates the considered graphs. At the next order, we rewrite
\begin{equation}
\quarter\sum_{j<k=1}^n \rho_k^{ab} \rho_j^{cd}s_{ab}s_{cd}
=
\sfrac{1}{8} \sum_{j<k=1}^n \rho_k^{ab} \rho_j^{cd}\comm{s_{ab}}{s_{cd}}
 +\sfrac{1}{8}\sum_{j,k=1}^n \rho_k^{ab} \rho_j^{cd}s_{ab}s_{cd}
 -\sfrac{1}{8} \sum_{k=1}^n \rho_k^{ab} \rho_k^{cd}s_{ab}s_{cd}.
\end{equation}
The first term on the right hand side reproduces the bi-local piece of the level-one generator. The second term is the product of two level-zero generators, which annihilates the diagrams under consideration and can thus be dropped. Noting that 
\begin{equation}
\rho_k^{ab} \rho_k^{cd}s_{ab}s_{cd}=(\Delta_k-4)\Delta_k \idop - 4\rho_k^{ab}s_{ab},
\end{equation}
 we can rewrite the last term according to
\begin{equation}
-\sfrac{1}{8}\sum_{k=1}^n\rho_k^{ab} \rho_k^{cd}s_{ab}s_{cd}=\sfrac{1}{8}\sum_{k=1}^n(4-\Delta_k)\Delta_k  \idop  +\half\sum_{k=1}^n \rho_k^{ab}s_{ab}.
\end{equation}
Here the first term cancels the piece proportional to the identity in \eqref{eq:moneqexp}, while the last term is the level-zero generator \eqref{eq:levzfrommon} and can thus be dropped. Collecting the remaining terms at order $u^{n-2}$ of \eqref{eq:moneqexp}, we thus find the level-one generator to be given by
\begin{equation}\label{eq:levofrommon}
\gen{\widehat J}^{ab}=
\sfrac{1}{8}\, f_{cd,ef}{}^{ab} \sum_{j<k=1}^n \rho_k^{cd} \rho_j^{ef}
+
 \half \sum_{k=1}^n v_k \, \rho_k^{ab}.
\end{equation}
Here $f_{cd,ef}{}^{ab}$ denotes the structure constants with $\comm{s_{ab}}{s_{cd}}=f_{ab,cd}{}^{ef} s_{ef}$ and
the evaluation parameters take the form
\begin{equation}
v_k=\half \mathop{\sum_{j=1}}_{j\neq k}^n\hat \delta_j.
\end{equation}
The level-zero and level-one generators in \eqref{eq:levzfrommon} and \eqref{eq:levofrommon}, respectively, agree with the expressions found on the single and double cross.

\paragraph{Cyclicity.}

An important point is the cyclicity of the level-one generators. In the case of the Yangian invariance of (cyclic) color-ordered tree-level scattering amplitudes in $\mathcal{N}=4$ SYM theory it is guaranteed due to a vanishing dual Coxeter number of the underlying symmetry algebra $\mathfrak{psu}(2,2|4)$ \cite{Drummond:2009fd}.  For the case at hand, i.e.\ for the conformal algebra $\alg{so}(2,4)$, the dual Coxeter number is non-vanishing. However, we still have some cyclicity constraints for certain types of fishnet graphs. For instance, an entire $n$-point fishnet graph (with complete rows only) has an $\sfrac{n}{2}$-site cyclic shift symmetry of the external legs. For this property to be compatible with the Yangian generators, we should thus wonder about the restrictions that this cyclicity imposes on the most generic level-one Yangian generator.

Consider the bi-local piece of the level-one generator (for convenience we use adjoint indices here)
\begin{equation}\label{eq:levobi}
\levo^A_\text{bi}|_{1n}=\half f^A{}_{BC} \sum_{j<k=1}^n \levz^C_j\levz^B_k,
\end{equation}
where $|_{1n}$ denotes the boundaries of summation.
Invariance of the considered graph under a cyclic shift by $m$ positions requires the following quantity to vanish:
\begin{equation}\label{eq:bicyc}
\levo^A_\text{bi}|_{1+m,n+m}-\levo^A_\text{bi}|_{1n}=\half f^A{}_{BC}f^{CB}{}_D \sum_{k=1}^m\levz^D_k-f^A{}_{BC}\sum_{k=1}^m\levz^C_k\levz^B.
\end{equation}
Restricting to the space of invariants under the level-zero symmetry, the second term is trivially zero. In order to compensate for the first term, we make use of the evaluation automorphism of the Yangian algebra and add the  contribution $\sum_{k=1}^n v_k \levz_k^A $ to the level-one generator \eqref{eq:levobi}.

The commutator of this additional term with an $m$-site cyclic shift yields the following contribution to \eqref{eq:bicyc}:\begin{equation}\label{eq:vdiffs}
\dots+\sum_{k=1}^n (v_k-v_{k-m})\levz_k^A.
\end{equation}
In order to compensate the first term in \eqref{eq:bicyc}, we should require that%
\footnote{
In order to make connection to the cyclicity statement in \eqref{cyclT}, we may alternatively compensate for the cyclic shift in \eqref{eq:levobi} by a modification of the evaluation parameters, i.e.\ by the replacement $v_k\to \tilde v_k$ for $k=1,\dots,n$. Hence, instead of the sum \eqref{eq:vdiffs} we add the term $\dots+\sum_{k=1}^n(v_k-\tilde v_k)\gen{J}_k^A$ to \eqref{eq:levobi}. For $m=-1$, this yields the conditions $\tilde v_k=v_k$ for $k=1,\dots,n-1$ and $\tilde v_n=v_n+\mathfrak{c}$. This modification of the last evaluation parameter corresponds to the shift of the inhomogeneity $u_n$ in \eqref{cyclT}.
}

\begin{equation}\label{eq:system}
v_k-v_{k-m}
=(\underbrace{\mathfrak{c}+a,\mathfrak{c}+a,\dots,\mathfrak{c}+a}_{m},\underbrace{a,\dots,a}_{n-m})_k.
\end{equation}
Here the constant $a$ attributes to the freedom to shift the level-one generator by a (complete) level-zero generator and $\mathfrak{c}$ denotes the dual Coxeter number defined via
\begin{equation}
f^A{}_{BC}f^{CB}{}_D=2\mathfrak{c}\,\delta^A_D .
\end{equation} 
Setting $m=\sfrac{n}{2}$, the above system of equations \eqref{eq:system} has $n+1$ free parameters and the solution reads
\begin{align}
v_{k>\frac{n}{2}}&=v_{k-\frac{n}{2}}-\frac{\mathfrak{c}}{2},
&
a&=-\frac{\mathfrak{c}}{2}.
\end{align}
Hence, in the case of entire fishnet graphs, half of the evaluation parameters $v_k$ are fixed by cyclicity.

\paragraph{Unique level-one generator for the box.}

We can apply the above cyclicity arguments to the single box integral alias the cross in coordinate space. The cross has a symmetry under single cyclic shifts of the external legs, i.e.\ it is invariant under the shift $x_k\to x_{k+m}$ for $m=1,2,3$. Setting $m=1,2,3$ in \eqref{eq:system}, the obtained system of equations is solved by
\begin{align}
v_k&=v-(k-1)\frac{\mathfrak{c}}{4},
&
a_m&=-m \frac{\mathfrak{c}}{4}.
\end{align}
Here $v$ represents the freedom to shift the level-one generator by a full level-zero generator and we may set this parameter to zero without loss of generality. Hence, for the single box integral the evaluation parameters $v_k$ are completely fixed by cyclicity. This means that for $n=4$ there is a unique choice for the level-one generator that is consistent with cyclicity.

\section{Dual Conformal Symmetry and the Yangian in Momentum Space}

In $\mathcal{N}=4$ SYM theory the Yangian invariance of scattering amplitudes is known to be equivalent to their superconformal and dual superconformal symmetry \cite{Drummond:2008vq,drummond2009yangian}. It is thus natural to ask whether a similar statement can be made rigorous in the case at hand --- namely, for planar scattering amplitudes in $\chi$FT$_4$ which enjoy ordinary conformal and dual conformal symmetry, cf.\  \cite{Henn:2011xk} for a discussion of the one-loop box. At the same time one may wonder whether an analogue of the above construction of the coordinate-space Yangian can also be performed in momentum space.  In what follows, we will explicitly demonstrate that the generator of special conformal transformations in the dual (coordinate) space can be rewritten as a Yangian level-one generator acting in on-shell momentum space. We thus derive the Yangian symmetry in momentum space and establish its equivalence to dual conformal symmetry.

\paragraph{Dual conformal symmetry.}
We start this section by briefly discussing the dual conformal symmetry of planar bi-scalar scattering amplitudes.%
\footnote{We are grateful to Jan Plefka for discussions on this point.}
To expose this symmetry it is convenient to pass to the dual variables which are defined through the relation 
\begin{align}
 p^{\mu}_{i} =x^{\mu}_i - x^{\mu}_{i+1} \, .
\label{eqn:dualmomenta}
\end{align}
In contrast to the situation in $\mathcal{N}=4$ SYM theory the dual conformal symmetry of the considered $\chi$FT$_4$ model is less universal in the sense that the dual conformal symmetry generators depend, at least in part, not only on the number of external legs but also on the structure of the amplitude itself. To make this statement more clear let us consider three simple examples of amplitudes:
\begin{figure}[H]
\centering
\includegraphicsbox[scale = 1]{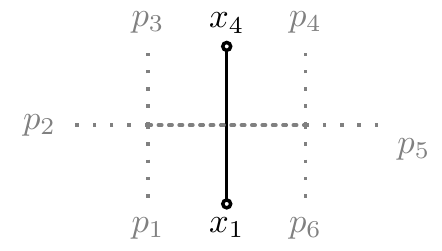}
\hfill
\includegraphicsbox[scale = 1]{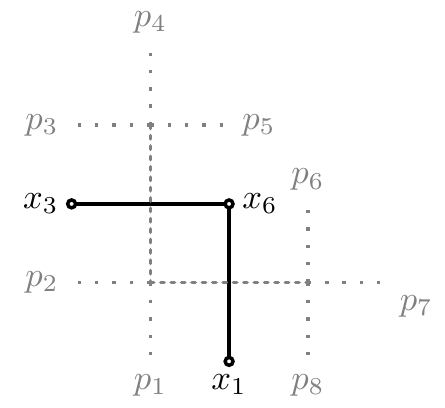}
\hfill
\includegraphicsbox[scale = 1]{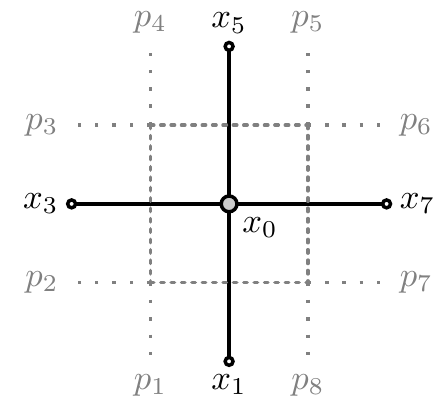}
\caption{Momentum space representation and dual space representation of three simple amplitudes in bi-scalar $\chi$FT$_4$. All momenta are taken to be ingoing.}
\end{figure}

Using Feynman rules we write down the following expressions for the amplitudes depicted above:
\begin{align}
A^{\mbox{\scriptsize{t}}}_6=\frac{\delta^{(4)}(x_1-x_7)}{x_{14}^2}\, , \hspace{1cm} A^{\mbox{\scriptsize{t}}}_{8}=\frac{\delta^{(4)}(x_1-x_9)}{x_{16}^2 x_{36}^2}\, , \hspace{1cm}
A^{\mbox{\scriptsize{1l}}}_{8}=\int  \mathrm{d}^4 x_{0} \:\: \frac{\delta^{(4)}(x_1-x_9)}{x_{10}^2x_{30}^2x_{50}^2x_{70}^2} \, , 
\label{eqn:threeamplitudes}
\end{align}   
A few comments concerning these amplitudes are in order. In equation \eqref{eqn:threeamplitudes} we have already introduced the dual coordinates \(x_1,x_2,\dots,x_n\) which are related to the external momenta as stated in equation \eqref{eqn:dualmomenta}. Obviously, the denominators just represent the region momenta flowing through the different propagators. Furthermore, note that in the above formulas we have relaxed the cyclicity condition on the $x$'s at the cost of a four-dimensional delta-function $\delta^{(4)}(x_1-x_{n+1})$ reimposing the closure of the light-like polygon. This delta-function corresponds to the momentum conserving delta-function $\delta^{(4)}(\genosr{P})$, where  \(\genosr{P}=\sum_{j=1}^n p_j\), and we have inserted it here for later convenience. On the contrary, we have not included the one-dimensional delta-functions ensuring the light-likeness of the edges $x_i-x_{i+1}$. The reason for this is that the dual conformal generators as well as the level-one on-shell momentum-space generators manifestly respect the light-likeness condition, so that we can safely disregard these delta-functions. This being said, let us now take a closer look at the dual conformal properties of these amplitudes. The representation of the dual conformal algebra that we will use here was introduced in equation \eqref{eqn:diffrepconfgen} and for pedagogical reasons we will start with a representation with conformal dimension $\Delta_i=0$ at each site. The amplitudes \eqref{eqn:threeamplitudes} are manifestly invariant under translations and rotations and thus they are annihilated by the corresponding generators. Acting with the dilatation generator on the amplitudes in equation \eqref{eqn:threeamplitudes} yields
\begin{align}
\gen{D} A^{\mbox{\scriptsize{t}}}_6= 6i A^{\mbox{\scriptsize{t}}}_6 \, , \hspace{1cm} \gen{D} A^{\mbox{\scriptsize{t}}}_8= 8i A^{\mbox{\scriptsize{t}}}_8 \, , \hspace{1cm} \gen{D} A^{\mbox{\scriptsize{1l}}}_{8}=8i  A^{\mbox{\scriptsize{1l}}}_{8} \, ,
\label{eqn:DonthreeAmps}
\end{align}
where 
\begin{align}
\gen{D}=-i \sum_{i=1}^{n+1} x_i^\mu \partial_{i \mu} \, .
\end{align}
Note that due to the relaxed cyclicity condition, the dual conformal generators are summed up to $n+1$ instead of $n$. From equation \eqref{eqn:DonthreeAmps} we anticipate that acting with the conformal dilatation generator on an amplitude just yields the number of external legs times $i$ and it is actually not too hard to convince oneself that this statement holds true for all the planar scattering amplitudes in $\chi$FT$_4$.
In order to make the dilatation generator a true symmetry generator we will now use the freedom to choose the conformal weights $\Delta_i$ to compensate for the terms on the right hand side of equation \eqref{eqn:DonthreeAmps}. The conformal weights $\Delta_i$ enter the dilatation generator in the following way:
\begin{align}
\gen{D'}=-i \sum_{i=1}^{n+1} \bigl( x_i^\mu \partial_{i \mu} + \Delta_i \bigr) \, .
\label{eqn:Dprime}
\end{align}
There are actually many choices that lead to a modified generator $\gen{D'}$ which annihilates the amplitudes \eqref{eqn:threeamplitudes} but the most natural one is:
\begin{align}
&\vec{\Delta}_{A^{\mbox{\scriptsize{t}}}_6}=(4+1,0,0,1,0,0,0) \, , \nonumber \\
&\vec{\Delta}_{A^{\mbox{\scriptsize{t}}}_8}=(4+1,0,1,0,0,2,0,0,0) \, , \nonumber \\
&\vec{\Delta}_{A^{\mbox{\scriptsize{1l}}}_8}=(4+1,0,1,0,1,0,1,0,0) \, .
\label{eqn:vecDelathreeexamp}
\end{align}
Note that we can always choose $\Delta_{n+1}=0$ as the delta-function allows us to eliminate the coordinate $x_{n+1}$ in favour of $x_1$. The factors of four in the above equations compensate for the weight that is introduced by the delta-function while all the other numbers are chosen such that the weight coming from the corresponding coordinate is cancelled out. Having discussed the dilatation symmetry, let us focus on special conformal transformations. Special conformal transformations are most conveniently studied by noting that the generator of special conformal transformations is related to the generator of translations by the following formula:
\begin{align}
\gen{K}^\mu= - \gen{I} \, \gen{P}^\mu \, \gen{I} \, .
\label{eqn:KviaInv}
\end{align}
Here, $\gen{I}$ is the inversion element which acts on the coordinates as $\gen{I}[x^\mu]=x^\mu/x^2$.
As a first step towards establishing the action of the generator $\gen{K}^\mu$ on the amplitudes \eqref{eqn:threeamplitudes} let us study their inversion properties. Using that the region momenta invert as
\begin{align}
\gen{I}[x_{ij}^2]=\frac{x_{ij}^2}{x_i^2 x_j^2} \, ,
\end{align}
as well as the formula $\gen{I}[\delta(x_1-x_{n+1})]=x_1^8 \delta(x_1-x_{n+1})$,\! \footnote{This formula can easily be derived by considering the definition of the delta-function $\int \mathrm{d}^4 x_1 \: \delta(x_1-x_{n+1})=1$ and noting that under an inversion the measure transforms as $\gen{I}[ \mathrm{d}^4 x_1]= \mathrm{d}^4 x_1/x_1^8$.} we find
\begin{align}
\gen{I}[A^{\mbox{\scriptsize{t}}}_6]= x_1^{10} x_4^2  A^{\mbox{\scriptsize{t}}}_6 \, , \hspace{1cm} \gen{I}[A^{\mbox{\scriptsize{t}}}_8]= x_1^{10} x_3^2 x_6^4  A^{\mbox{\scriptsize{t}}}_8 \, , \hspace{1cm} \gen{I}[A^{\mbox{\scriptsize{1l}}}_{8}]= x_1^{10} x_3^2 x_5^2 x_7^2  A^{\mbox{\scriptsize{1l}}}_{8} \, .
\end{align}
In contrast to the situation in $\mathcal{N}=4$ SYM theory the amplitudes obviously do not transform in a completely covariant way. This is on the one hand due to the fact that there is no supermomentum conserving delta-function present, which could balance out the inversion weight of the bosonic delta-function. On the other hand, also the amplitude functions themselves do not transform in a completely homogeneous way, as some of the $x$'s are simply not present while others come with a power higher than $2$.  Having studied the inversion properties of the amplitudes, we can now easily write down expressions for the action of the generator $\gen{K}^\mu$ on the three amplitudes \eqref{eqn:threeamplitudes}. Using equation \eqref{eqn:KviaInv} we find:
\begin{align}
\label{eqn:KonthreeA1}
&\gen{K}^{\da \a} A^{\mbox{\scriptsize{t}}}_6 = i \bigl(5 x_1^{\da \a} +  x_4^{\da \a}  \bigr) A^{\mbox{\scriptsize{t}}}_6 \, , \\
&\gen{K}^{\da \a} A^{\mbox{\scriptsize{t}}}_8 = i \bigl(5 x_1^{\da \a} + x_3^{\da \a} + 2 x_6^{\da \a}  \bigr) A^{\mbox{\scriptsize{t}}}_8 \, ,\\
&\gen{K}^{\da \a} A^{\mbox{\scriptsize{1l}}}_{8} = i \bigl(5 x_1^{\da \a} + x_3^{\da \a} +  x_5^{\da \a}  +  x_7^{\da \a} \bigr) A^{\mbox{\scriptsize{1l}}}_8 \, .
\label{eqn:KonthreeA3}
\end{align}
where we employed the standard definition:  for later convenience we have written \(x^\mu\) and  $\gen{K}^{\mu}$ as  $2\times2$ matrices being defined as
\begin{align}
&x^{\dot \alpha\alpha}= \bar{\sigma}_\mu^{\da \a} \gen{x}^\mu,\\
&\gen{K}^{\da \a}= \tfrac{1}{2} \bar{\sigma}_\mu^{\da \a} \gen{K}^\mu=-i \sum\limits_{i=1}^{n+1} x_i^{\da \b} x_i^{\db \a} \partial_{i \b \db} \, .
\end{align}
As in the case of dilatation symmetry we can now define a modified generator $\gen{K'}^{\da \a}$ which annihilates the amplitudes by setting
\begin{align}
\gen{K'}^{\da \a}= \gen{K}^{\da \a} - i \sum\limits_{i=1}^n \Delta_i \, x_i^{\da \a} \, ,
\label{eqn:defKprimethreeexamp}
\end{align}
with the $\Delta_i$'s as defined in equation \eqref{eqn:vecDelathreeexamp}. Our examples and equation \eqref{eqn:vecDelathreeexamp} make it clear that the generators $\gen{D'}$ and $\gen{K'}^{\da \a}$ are no longer universal as the vector $\vec{\Delta}$ does not only depend on the number of external legs but also on the amplitude itself. However, note that the generators $\gen{D'}$ and $\gen{K'}^{\da \a}$ are perfectly consistent with the algebraic restrictions imposed on them by the conformal commutation relations. Hence, the generators $\gen{P}_\mu$, $\gen{L}_{\mu \nu}$, $\gen{D}'$ and $\gen{K}'_\mu$  still furnish a representation of the conformal algebra $\mathfrak{so}(2,4)$. \\

Finally, let us comment on the situation for a generic planar amplitude in bi-scalar $\chi$FT$_4$. As already mentioned above, the plain dilatation generator $\gen{D}$ acts on an amplitude as follows:
\begin{align}
\gen{D} \, A_n =i n \, A_n \, ,
\label{eqn:DonA}
\end{align}
For a given planar amplitude the modified dilatation generator 
\begin{align}
\gen{D'}=-i \sum_{i=1}^{n+1} \bigl( x_i^\mu \partial_{i \mu} + \Delta_i \bigr) \, ,
\end{align}
annihilating the amplitude, can be constructed in the following way: first, we determine which $x_i$'s will be absent in the amplitude by drawing the amplitudes' dual graph and setting to zero all the corresponding $\Delta_i$'s. For the remaining $x_i$'s we set the corresponding $\Delta_i$'s equal to the number of lines which meet in the point $x_i$. Finally, we set to zero $\Delta_{n+1}$ and add a factor of $4$ to $\Delta_1$ to compensate for the weight of the delta-function. The resulting generator $\gen{D'}$ will then annihilate the considered amplitude. The generator of special conformal transformations annihilating this amplitude follows immediately from the algebra. Explicitly, it reads:
\begin{align}
\gen{K'}^{\da \a}= \gen{K}^{\da \a} - i \sum\limits_{i=1}^n \Delta_i \, x_i^{\da \a} \, .
\label{eqn:defKprime}
\end{align}
Finally, note that due to equation \eqref{eqn:DonA}, the $\Delta_i$'s satisfy the following relation:
\begin{align}
\sum\limits_{i=1}^n \Delta_i=n \, .
\label{eqn:alphaconstraint}
\end{align}
We will use this equation in the next paragraph, when we establish the connection between the generator $\gen{K'}^{\da \a}$ and the level-one momentum generator.

\paragraph{Yangian symmetry in momentum space.}  
In this paragraph we will now demonstrate that the generator $\gen{K'}^{\da \a}$ agrees with the conformal level-one momentum generator up to terms which annihilate the amplitudes by themselves. The discussion follows closely the one presented in \cite{drummond2009yangian}, where the statement was proved for the case of $\mathfrak{psu}(2,2|4)$. To rewrite $\gen{K'}$ as an operator acting in on-shell spinor helicity space, we first extend it such that it commutes with the constraint 
\begin{align}
x_i^{\dg \g} - x_{i+1}^{\dg \g} - \la^\g_i \tla^\dg_i=0 \, .
\label{eqn:constrainglambdax}
\end{align}
The result reads
\begin{align}
\gen{K'}^{\da \a}=-i \Biggl[ \sum\limits_{i=1}^{n+1} x_i^{\da \b} x_i^{\db \a} \partial_{i \b \db} + \sum\limits_{i=1}^{n} \bigl( x_i^{\da \b} \la_i^\a \partial_{i \b} + x_{i+1}^{\db \a} \tla_i^\da \tilde{\partial}_{i \db} + \Delta_i \, x_i^{\da \a} \bigr) \Biggr] \, .
\end{align}
Using the inverse of equation \eqref{eqn:constrainglambdax}
\begin{align}
x^{\da \a}_i=x^{\da \a}_1 - \sum\limits_{j < i=1}^n \tla_j^\da \la_j^\a 
\end{align}
and dropping the term that includes a derivative with respect to $x$, we find
\begin{align}
\gen{K'}^{\da \a}=& i \sum\limits_{j<i=1}^n \bigl(\tla_j^\da \la_j^\b   \la_i^\a \partial_{i \b} + \Delta_i \,  \tla_j^\da \la_j^\a \bigr)+ i \sum\limits_{j<i+1=1}^n \bigl(\tla_j^\db \la_j^\a \tla_i^\da \tilde{\partial}_{i \db}\bigr) \nonumber \\
&-i \sum\limits_{i=1}^n \bigl( x^{\da \b}_1 \la_i^\a \partial_{i \b} +x_1^{\db \a} \tla_i^\da \tilde{\partial}_{i \db} +  \Delta_i \, x^{\da \a}_1\bigr) \, .
\label{eqn:Kprimeonshell}
\end{align}
Note that the amplitudes can always be written as distributions depending exclusively on the spinor helicity variables $\{\la_i\}$ and $\{\tla_i\}$. For this reason we could safely disregard all terms containing a derivative with respect to the dual coordinates. Starting from equation \eqref{eqn:Kprimeonshell} it is now however a straightforward exercise to rewrite $\gen{K'}^{\da \a}$ as
\begin{align}
\gen{K'}^{\da \a}=&\tfrac{i}{2} \genosr{\widehat{P}}^{\da \a}_{\mathrm{bi}} + i \sum\limits_{j<i=1}^n (\Delta_i-1) \genosr{P}_j^{\da \a} + \tfrac{i}{2} \bigl(\genosr{P}^{ \da \b} \genosr{L}^{\a}{}_{\b} + \genosr{P}^{\db \a } \genosr{\bar{L}}^{\da}{}_{\db}  +\genosr{P}^{\da \a} \genosr{D} - \genosr{P}^{\da \a} \bigr)\nonumber \\
&-\tfrac{i}{2} \sum\limits_{i=1}^n \genosr{P}_i^{\da \a} \bigl( \la_i^\g \partial_{i \g}-\tla_i^\dg \partial_{i \dg}\bigr) -i \sum\limits_{i=1}^n \bigl(x^{\da \b}_1 \genosr{L}^{\a}_{i}{}_{\b}+x^{ \db \a}_1 \genosr{\bar{L}}^{\da}_{i}{}_{\db}+x^{\da \a}_1 \genosr{D}_i \bigr) \, ,
\label{eqn:KaslevoPinterm}
\end{align}
where we have introduced the conformal generators written in terms of spinor helicity variables
\begin{align}
\genosr{L}^{\a}_{i}{}_{\b}&=\lambda^{\a}_i  \partial_{ i \b} - \sfrac{1}{2} \delta^{\a}_{\b} \lambda^{\g}_i  \partial_{i \g}\, ,  \hspace{1,4cm} \bar{\genosr{L}}^{\da}_{i}{}_{\db}=\tilde\lambda^{\da}_i  \tilde{\partial}_{i \db}-\sfrac{1}{2} \delta^{\da}_{\db}\tilde\lambda^{\dg}_{i}  \tilde{\partial}_{ i\dg} \, ,\nonumber \\
\genosr{P}_i^{\da \a}&= \tilde\lambda^{\da}_i  \lambda^{\a}_i \, , \hspace{3,7cm}
\genosr{D}_i=\sfrac{1}{2}  \lambda^\a_i  \partial_{i \a} + \sfrac{1}{2}  \tilde\lambda^\da_i  \tilde{ \partial}_{i \da}+1 \, .
\label{eqn:confgenspinor}
\end{align}
The generator $\genosr{\widehat{P}}^{\da \a}_{\mathrm{bi}}$ in equation \eqref{eqn:KaslevoPinterm} is the level-one momentum generator
\begin{align}
\genosr{\widehat{P}}^{\da \a}_{\mathrm{bi}}=\sum\limits_{j<i=1}^n \bigl(\genosr{P}^{\da \b}_j \genosr{L}^{\a}_{i}{}_{\b} + \genosr{P}^{\db \a}_j \genosr{\bar{L}}^{\da}_{i}{}_{\db}  +\genosr{P}^{\da \a}_{j} \genosr{D}_{i} - (i \leftrightarrow j ) \bigr) \, ,
\end{align}
as it follows from the formula \eqref{eqn:generallevone} with the underlying level-zero algebra being the conformal algebra spanned by the generators \eqref{eqn:confgenspinor}. Note that in order to bring $\gen{K'}^{\da \a}$ to the above mentioned form, we have also used the constraint equation \eqref{eqn:alphaconstraint}, which allowed us to replace the $\Delta_i$'s by $1$ in the term that contributes to $x^{\da \a}_1 \genosr{D}_i$.
Finally, using the level-zero invariance of the amplitudes as well as the fact that all the external particles have zero helicity, i.e.\
\begin{align}
\tfrac{1}{2} \bigl( \la_i^\g \partial_{i \g}-\tla_i^\dg \partial_{i \dg}\bigr) A_n=0 \, ,
\end{align}
we see that most of the terms on the right hand side of equation \eqref{eqn:KaslevoPinterm} drop out, leaving us with
\begin{align}
\gen{K'}^{\da \a}=&\tfrac{i}{2} \genosr{\widehat{P}}^{\da \a}_{\mathrm{bi}} + i \sum\limits_{j<i=1}^n (\Delta_i-1) \genosr{P}_j^{\da \a} \, .
\end{align}
The local term would obviously vanish if all the $\Delta_i$'s were equal to $1$ as they are for example in the case of $\mathcal{N}=4$ SYM theory. However, since this is not the case here, we arrive at a purely bosonic Yangian generator with non-vanishing evaluation parameters, which is in complete analogy with the $x$-space level-one momentum generator that we considered before.

\section{Discussion and Outlook}

In this paper we established the conformal Yangian symmetry of single-trace correlators and amplitudes in the planar approximation  of the bi-scalar  $\chi\text{FT}_4$  theory \eqref{Lthree}. This theory appeared in a specific, double scaling limit of \(\gamma\)-twisted  \({\cal\ N}=4\) SYM theory  \cite{Gurdogan:2015csr}.   Each of 
the above observables is given by a single, generically multi-loop Feynman graph with the topology of a disc.  In the bulk, the disc typically has the  structure of  Zamolodchikov's fishnet Feynman graphs~\cite{Zamolodchikov:1980mb}, i.e.\ it represents a piece of regular square lattice.   The infinite-dimensional Yangian over the conformal algebra $\alg{so}(2,4)$ was explicitly constructed in its RTT realization. Here, the above coordinate-space Feynman graphs were shown to furnish eigenstates of an inhomogeneous monodromy matrix in the spirit of the work \cite{Chicherin:2013sqa,Chicherin:2013ora} (see also \cite{Frassek:2013xza}).   The  \(\alg{so}(2,4)\) Lax operators forming this monodromy have specific inhomogeneity parameters  depending on the shape of the boundary.  

Via expansion of this monodromy we obtained the respective level-zero Lie algebra generators and the bi-local level-one generators of the Yangian in its first realization, which annihilate the expressions represented by fishnet diagrams. For graphs with massless external legs, alias massless scattering amplitudes in the bi-scalar theory, we then demonstrated that the dual conformal symmetry is equivalent to a Yangian level-one symmetry in momentum space.

Importantly, the above Feynman integrals can be argued to be free of divergencies, and hence there is no need for introducing a regulator which could break the conformal (Yangian) symmetry. As opposed to $\mathcal{N}=4$ SYM theory, the Yangian symmetry of the full (all-loop) planar scattering matrix of the bi-scalar $\chi\text{FT}_4$ is thus an exact statement.
Moreover, the breakdown of conformal symmetry by the double-trace terms in the action 
 \cite{Fokken2014a,Fokken2014b,Fokken2014c,Sieg:2016vap}
seems to be not an issue here since we simply do not have such anomalous amplitudes in the planar limit.

The Yangian provides bi-local (with respect to the   coordinates of  external legs)  differential equations for all Feynman integrals of fishnet type with the disc topology. Notably, at present only the simplest of these integrals, i.e.\ the one-loop scalar box, has been solved. We are optimistic that the discovered Yangian symmetry will open the door to computing the respective higher-loop integrals via the powerful toolbox of integrability, as it happened to large classes of multi-loop graphs of bi-scalar  $\chi\text{FT}_4$  theory, such as ``wheel''-graphs \cite{Gurdogan:2015csr,GKKNS} and magnon correlator graphs \cite{Caetano:2016ydc} relevant for the computation of anomalous dimensions. 

Taking the bi-scalar $\chi\text{FT}_4$ as a starting point, we may wonder what the above precise formulation of its integrability teaches us about $\mathcal{N}=4$ SYM theory. In particular, it would be interesting to look for a connection to the Yangian symmetry that lurks behind the Qbar-equation and its level-one counterpart for the finite BDS-subtracted S-matrix of $\mathcal{N}=4$ SYM theory \cite{CaronHuot:2011kk}. As a starting point, one might try to understand the symmetries of scattering amplitudes in the $\gamma$-deformed $\mathcal{N}=4$ SYM theory as an expansion around the bi-scalar case considered here.

Recently, the Yangian symmetry of $\mathcal{N}=4$ super Yang--Mills theory was understood on the level of its action \cite{Beisert:2017pnr}. It would be highly interesting to adapt the developed criterion for the integrability of planar gauge theories in four dimensions to the bi-scalar theory under investigation. Eventually, this might allow to derive our Yangian symmetry of correlators and amplitudes from the Lagrangian.

As is well known, massless scattering amplitudes in $\mathcal{N}=4$ SYM theory are dual to polygonal Wilson loops with light-like edges in the strong-coupling \cite{Alday:2007hr} and weak-coupling \cite{Drummond:2007aua} regimes. This duality serves as a natural explanation of the ordinary and dual conformal, alias Yangian symmetry of scattering amplitudes. Due to the absence of a gauge field in the bi-scalar $\chi\text{FT}_4$, the definition of a Wilson loop and hence a possible translation of this duality is not obvious. In order to better understand how to formulate a Wilson loop in this theory, it may be fruitful to first forget about the polygonal contour and to consider the bi-scalar limit of a $\gamma$-deformed smooth Maldacena--Wilson loop \cite{Maldacena:1998im,Rey:1998ik} which, in addition to the gauge field, also couples to the scalars of the theory.

Let us recall that every Yangian-invariant scattering amplitude in $\mathcal{N}=4$ SYM theory can be written as an integral over a Grassmannian
\cite{ArkaniHamed:2012nw}. Identifying a similar geometric structure for the bi-scalar amplitudes at hand would certainly be of great importance.

Note that the considered scalar amplitudes do not exhaust all possible amplitudes of the bi-scalar theory, not even in the planar limit. We can also include external states described by derivatives of scalar fields and various boundary OPE's of such operators. It would be interesting to understand whether the Yangian symmetry extends to such amplitudes and single-trace correlators.

Finally, a similar Yangian symmetry of planar  amplitudes exists in the three-dimensional analogue of the bi-scalar theory, which can be obtained by a similar double scaling limit of the three-dimensional \(\gamma\)-deformed ABJM model (``tri-scalar'' theory)  \cite{Caetano:2016ydc}, where we deal with regular triangular fishnet graphs.
The same holds for a similar six-dimensional tri-scalar theory with chiral cubic interactions, recently studied in~\cite{Mamroud:2017uyz}, and dominated by regular hexagonal fishnet graphs (these models exhaust all three types of graphs whose integrability was noticed by A.~Zamolodchikov).\footnote{Both 3D and 6D theories  appear to be  true CFT's in the planar limit~\cite{Mamroud:2017uyz}.} We will address these questions in future work.

\section*{Acknowledgments}
\label{sec:acknowledgments}

We  are  thankful to B.~Basso,  J.~Caetano, L.~Dixon, J.~Henn, G.~Korchemsky and J.~Plefka for discussions. 
The work of V.K.\ and D-l.Zh.\ was supported by the People Programme (Marie
Curie Actions) of the European Union's Seventh Framework Programme
FP7/2007-2013/ under REA Grant Agreement No.317089 (GATIS), by the
European Research Council (Programme ``Ideas'' ERC-2012-AdG 320769
AdS-CFT-solvable).
 V.K.\ is grateful to Humboldt University (Berlin) for the hospitality and
financial support of this work in the framework of the ``Kosmos'' programme. D.M.\ gratefully acknowledges the hospitality of the Mainz Institute for Theoretical Physics during the workshop ``Amplitudes: Practical and Theoretical Developments''.

\appendix

\section{Cyclicity}\label{AppCycl}

In this Appendix we prove the cyclicity property (\ref{cyclT}) of the eigenvalue problem adopting the arguments from \cite{Chicherin:2013sqa} to the conformal Lax (\ref{Lax}).

We use shorthand notations $L(u_{\pm}) \equiv L(u_+,u_-)$ and $u_{\pm} \equiv u_+ u_-$.
We need the inversion formula for the Lax and for its matrix transpose $L^t$,
\begin{align}
& L^{-1}(u_{\pm}) = - u_{\pm}^{-1} \,L(-u_{\mp}) \;, \label{inv}\\
& (L^t)^{-1}(u_{\pm}) = - (u_{\pm} + 2) ^{-1} \,L^t(-u_{\mp} - 4)  \label{trans}
\end{align} 

To prove the cyclicity we apply several times inversions and matrix transpositions:
 
Step 1. We start with the eigenvalue relation
\begin{align*}
L_n(u_{n\pm}) \ldots L_1(u_{1\pm}) \ket{G} = \lambda\, \ket{G} \idop
\end{align*}

Step 2. We invert $L_n$,
\begin{align*}
L_{n-1}(u_{n-1\pm}) \ldots L_1(u_{1\pm}) \ket{G} = \lambda\, L^{-1}_n(u_{n\pm})\ket{G}
\end{align*}

Step 3. We act by matrix transposition on both sides of the previous eq.,
\begin{align*}
L^t_1(u_{1\pm})  \ldots L^{t}_{n-1}(u_{n-1\pm}) \ket{G} = \lambda\, L^{-1 t}_n(u_{n\pm})\ket{G}
\end{align*}

Step 4. We again invert $L_n$ applying eqs. (\ref{inv}), (\ref{trans}),
\begin{align*}
L^{t}_n(u_{n\pm}-4) L^t_1(u_{1\pm})  \ldots L^{t}_{n-1}(u_{n-1\pm}) \ket{G} = \widetilde\lambda \ket{G} \idop
\end{align*}
where $\widetilde{\lambda} \equiv \lambda\cdot  u^{-1}_{n\pm} \cdot (u_{n\pm}-2)$.

Step 5. We act by matrix transposition on both sides of the previous eq.,
\begin{align*}
L_{n-1}(u_{n-1\pm}) \ldots L_1(u_{1\pm}) L_n(u_{n\pm}-4)\ket{G} = \widetilde{\lambda}\, \ket{G} \idop
\end{align*}

Eq. (\ref{cyclT}) is proven.

\section{Cross Integral}\label{AppB}

In this appendix we provide algebraic expressions corresponding to the sequence of monodromy contour transformations in Fig. \ref{crossfig}.

Step 1. We act with four-point monodromy onto cross integral (\ref{cross}),
\begin{align*}
\int \mathrm{d}^4 x_0 \,  L_4[4,5]L_3[3,4]L_2[2,3]L_1[1,2] x_{10}^{-2}x_{20}^{-2}x_{30}^{-2}x_{40}^{-2} \,.
\end{align*}

Step 2. We multiply the monodromy by an identity matrix $\idop = [2]^{-1}\, L_0^T[2,0]\cdot 1$ on the right hand side, eq. (\ref{transpose}), and integrate the inserted Lax by parts, i.e. $L_0^T \to L_0$,
\begin{align}
&\int \mathrm{d}^4 x_0 \, L_4[4,5]L_3[3,4]L_2[2,3]L_1[1,2](L_0^T[2,0]\cdot 1) \, x_{10}^{-2}x_{20}^{-2}x_{30}^{-2}x_{40}^{-2} \notag \\ = 
&\int \mathrm{d}^4 x_0 \, L_4[4,5]L_3[3,4]L_2[2,3]L_1[1,2]L_0[2,0] x_{10}^{-2} x_{20}^{-2}x_{30}^{-2}x_{40}^{-2}\,.
\end{align}
In the following steps we show that the integrand is an eigenfunction of the five-point monodromy,
\begin{align}
L_4[4,5]L_3[3,4]L_2[2,3]L_1[1,2]L_0[2,0]\, x_{10}^{-2} x_{20}^{-2}x_{30}^{-2}x_{40}^{-2} = 
[2][3][4]^2[5]\, x_{10}^{-2} x_{20}^{-2}x_{30}^{-2}x_{40}^{-2}\,.   
\label{integrand}
\end{align}

Step 3. 
We use intertwining relation\footnote{We highlight permuted parameters of the Laxes in the intertwining relation.} (\ref{intw}) 
for adjacent Laxes $L_1$ and $L_0$, and then we act by $L_1$ onto $1$ according to eq. (\ref{vacuum}),
\begin{align*}
L_1[{\bf 1},2] L_0[2,{\bf 0}]\, x_{10}^{-2} = x_{10}^{-2}\,\underbrace{L_1[{\bf 0},2]}_{[2]\idop} L_0[2,{\bf 1}]\,.
\end{align*}
So one Lax drops out of the monodromy and the five-point monodromy reduces to the four-point monodromy. 

Step 4. We repeat analogous simplification for adjacent Laxes $L_2$ and $L_0$
again reducing the length of the monodromy
\begin{align*}
L_2[{\bf 2},3]L_0[2,{\bf 1}]\,x_{20}^{-2} =
x_{20}^{-2} \, \underbrace{L_2[{\bf 1},3]}_{[3]\idop}L_0[2,{\bf 2}]\,.
\end{align*}

Step 5. We implement simplification for adjacent Laxes $L_3$ and $L_0$ 
\begin{align*}
L_3[{\bf 3},4]L_0[2,{\bf 2}]x_{30}^{-2} =
x_{30}^{-2} \,\underbrace{L_3[{\bf 2},4]}_{[4]\idop}L_0[2,{\bf 3}]\,.
\end{align*}

Step 6. Finally, we have the two-point monodromy and find that the propagator is its eigenfunction 
\begin{align*}
L_4[{\bf 4},5]L_0[2,{\bf 3}]x_{40}^{-2} =
x_{40}^{-2}\, L_3[{\bf 3},5] L_0[2,{\bf 4}] = x_{40}^{-2}\, [4][5]\,\idop\,.
\end{align*}
Collecting all numerical factors appeared in the previous steps we obtain eq. (\ref{integrand}) and eq. (\ref{crossef}).

\section{Expansion of Monodromy Eigenvalue}
\label{sec:eigenexp}
Here we give the first five orders of the spectral-parameter expansion of the monodromy eigenvalue function:
\begin{align}\label{eq:expansionRHSfirst5}
\lambda(\vec{u})=
u^n 
&+\half u^{n-1}  \sum_{k=1}^n \hat \delta_k
+\quarter u^{n-2} \Big[\sum_{i<j=1}^n \hat \delta_i \hat \delta_j -\sfrac{1}{2}\sum_{j=1}^n\hat \Delta_j  \Big]
\nonumber\\
&+\sfrac{1}{8}u^{n-3}\Big[
\sum_{i<j<k=1}^n\hat \delta_i\hat\delta_j\hat\delta_k
-\half\sum_{j,k}P_{jk}\hat \Delta_j\hat\delta_k
\Big]
\nonumber\\
&+\sfrac{1}{16}u^{n-4}\bigg[
\sum_{i<j<k<l=1}^n\hat \delta_i\hat\delta_j\hat\delta_k\hat\delta_l
-\quarter\sum_{j,k,l=1}^n
\big(1-P_{jl}\delta_{lk}-4P_{jl}\delta_{jk}-4\delta_{lk}\delta_{jl}
\big)\hat\Delta_j\hat\delta_k\hat\delta_l
\nonumber\\
&\qquad\qquad\quad+\sfrac{1}{8} \sum_{i,j=1}^n P_{ij} \hat \Delta_i \hat\Delta_j
+4\sum_{i=1}^n \hat \Delta_i 
\bigg]
+\mathcal{O}(u^{n-5}).
\end{align}
Here we use the shorthand notations $\hat \delta_k=\delta_k^++\delta_k^-+2$ and $\hat\Delta_i=\Delta_i(\Delta_i-4) $ with  $\Delta_k=\delta_k^+-\delta_k^-+2$ and we set $P_{jk}=1-2\delta_{jk}$.

\bibliographystyle{JHEP}
\bibliography{publications}

\end{document}